# Water Ice Compression: Principles and Applications

Lei Li[1], Maolin Bo[1], Jibiao Li[1], Xi Zhang[2], Yongli Huang[3,*], Chang Q Sun[4,*]

**Abstract**

A set of concepts cultivated in the past decade has substantiated the understanding of water ice compression. The concepts include the coupled hydrogen bond (O:H−O) segmental cooperativity, segmental specific-heat disparity, compressive O:H−O symmetrization and polarization, and quasisolidity. Computational and spectrometric outcome consistently justifies that the intermolecular H−O bond has a negative compressibility while the O:H nonbond does contrastingly because of the involvement of the interionic O—O repulsive coupling. Compression lowers the melting point $T_m$ (named regelation) and raises the freezing temperature $T_N$ (for instant ice formation) by inward shifting the quasisolid phase boundary through Einstein's relation, $\Delta\Theta_{Dx} \propto \Delta\omega_x$, which is in contrast to the effects of electrostatic polarization and water molecular undercoordination. Theoretical reproduction of the phase boundaries in the phase diagram revealed that H−O bond relaxation dictates boundaries of negative slope, $dT_C/dP < 0$, and that the O:H relaxation determines those of $dT_C/dP > 0$. O:H−O bond frozen governs boundaries of constant $T_C$ such as the $I_C$-XI boundary and the H−O and O:H energy compensation yields those of constant $P_C$ such as X-VII/VIII boundaries. The O—O repulsion opposes compression minimizing the compressibility. Polarization enlarges the bandgap and the dielectric permittivity of water ice by raising the nonbonding states above the Fermi energy. Progress evidences the efficiency and essentiality of the coupled O:H−O bonding and electronic dynamics in revealing the core physics and chemistry of water ice, which could extend to other molecular crystals such as energetic materials.

Keywords: hydrogen bond; specific-heat; regelation; phase transition; bandgap; compressibility

Contents

[1] EBEAM, School of Materials Science and Engineering, Yangtze Normal University, Chongqing 408100, China (Lilei@yznu.edu.cn; bmlw@yznu.edu.cn; jibiaoli@yznu.edu.cn)
[2] Institute of Nanosurface Science and Engineering, Shenzhen University, Shenzhen 518060, China (zh0005xi@szu.edu.cn)
[3] Key Laboratory of Low-dimensional Materials and Application Technology, School of Materials Science and Engineering, Xiangtan University, Xiangtan 411105, China (Huangyongli@xtu.edu.cn)
[4] School of EEE, Nanyang Technological University, Singapore 639798 (ecqsun@ntu.edu.sg; corresponding author)

# 1 Wonders of water ice compression

*− Pressure derives anomalies by mediating the hydrogen bonding and electronic dynamics*

Anomalies of water ice derived by mechanical compression ($P$) have attracted much attention since 1859 when Faraday and Thomson discovered the phenomenon of Regelation [1, 2] – ice melts under compression and freezes again when the pressure is relieved with mechanisms that reman debating. One can cut a block of ice through using a loaded wire without breaking the block [3]. The wire's track is refilled as soon as the pressure is relieved, so the ice block will remain a solid after the wire passes completely through [4-6]. The phenomenon of Regelation enables liquid water flow from the base of a glacier to lower elevations, as the glacier exerts a pressure that is sufficient to melt ice of its lower surface [7]. A 210 MPa pressure can lower the melting temperature $T_m$ to 251 °K. Conversely, a negative -95 MPa pressure or tension raises the $T_m$ by 6.5 °K [8], which shows why a water pipe does not bluster at temperature 250 K in the winter time.

The Regelation is accompanied with elevation of the critical temperatures ($T_C$) for ice nucleation (freezing) $T_N$ and vaporization (boiling) $T_V$ [9]. Pressure below the atmospheric value at a high latitude raises the $T_m$ but lowers the $T_N$ and $T_V$, making water boiling easier at temperature below 373 °K, as one can observe. Another example is that one can make "instant ice" by slamming (supplying a pressure impulse to raise the $T_N$) a bottle of supercooled water taken from fridge. Likewise, increasing the pressure from 1 to 50 GPa can lower the $T_C$ from 280 to 150 °K for the ice VII–VIII phase transition [10-12]. Therefore, the $T_C$ for a phase boundary may go up or down or keep constant with pressure.

The O:H−O bond is asymmetric of the ~0.17 and ~0.10 nm segmental length in the Ice–VIII phase. The O:H and H−O length asymmetry turns to be symmetry of identical 0.11 nm length when enters the X phase under compression. In the X phase, the $H^+$ proton lies midway between adjacent $O^{2-}$ ions [13, 14]. The proton centralization occurs to ice under 60 GPa pressure and 85-100 °K temperature [15, 16]. In contrast, the proton centralization happens to liquid $D_2O$ at 70 GPa pressure and 300 °K temperature [17]. Investigations showed that the superionic $2H_2O \leftrightarrow H_3O^+ : HO^-$ transition in which every fourth O:H−O bond reverses its direction takes place at 0.1 – 2.0 TPa pressure and 2000 °K temperature [18, 19]. In the supercritical states of $T \geq 637$ °K and $P \geq 22.1$ MPa, water molecules perform like compressed monomers with the H−O bond vibrating in the same frequency of 3610 cm$^{-1}$ for the H−O dangling bond of water surface [20]. Therefore, the O:H-O bond performs differently at the phase boundaries and in regimes of pressures and temperatures.

The phonon frequency ω of a specific vibration mode features its corresponding bond length $d$ and energy $E$, in the form of $\omega^2 \propto E/d^2$. Compression stiffens all phonons of a 'normal' atomistic substance without an exception [21]. However, compression stiffens the softer O:H phonons ($\omega_L < 300$ cm$^{-1}$) and softens the stiffer H−O phonon ($\omega_H > 3000$ cm$^{-1}$) for most phases or water ice (subscript H and L represents the high and low frequencies) except for those under extreme conditions such as the X phase above 60 GPa and the supercritical states as afore mentioned. Phonon frequencies transit abruptly at phase boundaries, which serves as the identifier of a phase transition [11, 12, 22-24].

Quantum computations suggested that the O:H contributes positively while the H−O negatively to the lattice energy of water ice under compression up to 2 GPa pressure [25]. Alternatively, O:H gains by its contraction but H-O losses energy by its elongation under compression. Compression also widens the bandgap and raises the dielectric constant of ice. The refractive index of liquid water at room temperature increases with pressure in the same thermal trend of its mass density [26, 27]. Bandgap expansion results in observation of blue ice under compression [28]. Compression raises the drift diffusivity or lowers the viscosity of the liquid at temperatures around 306 °K temperature and 100 – 200 MPa pressure [29]. Liquid water reaches its lowest compressibility and maximal thermal diffusivity at temperatures around 320 °K [30]. Therefore, the anomalous O:H-O relaxation and its electronic dynamics modulate not only the structure order but also properties of water and ice.

The pressure derived anomalies of water ice are exceedingly interesting and useful [31]. However, understanding the anomalies of water ice compression stays an issue of challenging. The classical continuum thermodynamics deals with water ice as a collection of vapor-like neutral particles to examine the response of the entire body to the applied stimuli. One often related a detectable property $Q(PV, ST, \ldots)$ such as the compressibility, elasticity, thermal stability, and viscosity of water ice directly to the external stimuli such as pressure $P$, volume $V$, and temperature $T$. The classical scheme formulated the liquid-vapor phase transition in terms of enthalpy, entropy, and Gibbs free energy [32]. For instances, Clausius–Clapeyron [33] described water-vapor transition near the standard atmosphere temperature and pressure and August-Roche-Magnu [32] approximated the temperature dependence of the saturation vapor pressure, $P_s$, in the following forms:

$$\begin{cases} \dfrac{dP_s}{dT} = \dfrac{L_v(T)P_s}{R_v T^2} & (Clausius-Clapeyron) \\ P_s(T) = 6.1094 \exp\left(\dfrac{17.625T}{T+243.04}\right) & (August-Roche-Magnus) \end{cases}$$

(1)

Where $L_v$ is the specific latent heat of evaporation of water and $R_v$ is the gas constant of vapor.

Molecular dynamics (MD) [34-37] treats the flexible or rigid, polarizable or non-polarizable, individual molecule as the primary unit of structure of water ice. The TIPnP (n = 1 ~ 5) series and the TIP4Q/2005 model have been elegantly used to model the performance of water ice [38-40]. The combination of MD and ultrafast-IR spectroscopy reveal information on the performance of molecules in the spatial-temporal domains such the residing time of molecules in a specific coordination sites of different viscosity. A combination of MD calculations [41] and IR and Raman spectroscopy [16, 42-45] derived phase boundaries for the (VII, VII)–X phase transition. In the MD wise, the proton was assumed to hop or tunnel in a $10^{-12}$ sec period between the asymmetrical sites of adjacent $O^{2-}$ anions or in Pauling's "two in and two out" ice rule [46, 47]. The proton was ascribed as undergoing "translational quantum tunneling" between the equivalent potential wells [13, 48, 49] that merge into single one located midway between $O^{2-}$ ions under high pressure [50].

The recent developed premise of the coupled O:H-O bonding and electronic dynamics takes the O:H-O as the basic energetic-functional-structural unit to feature the performance of bonds, electrons, and molecules in the energetic-spatial-temporal domains. The coupled O:H-O bond encompasses the density functional theory, MD, Fourier fluid thermodynamics, Lagrangian oscillating dynamics, and the dynamic and static electron and phonon spectrometrics to reveal the details of the O:H-O relaxation and electron polarization and its derivatives. This set of strategies has enabled a resolution to a number of anomalies of water ice when subjecting to electrostatic polarization [51], compressing [52], heating [53], molecular undercoordination [54], solvation of acids, bases, salts, and organic molecular solutes . Interested readers are referred to recent work [55, 56].

How does water ice respond to pressure so strangely? How does compression matter the performance of water ice?

This presentation aims to show the essentiality of the "O:H-O bonding and electronic dynamics" in resolving anomalies demonstrated by water ice when responding to compression because the coordination bonding and electronic dynamics stem uniquely the performance of bonds, electrons, and molecules in the spatial-temporal-energic domains [57] and the structures and properties of molecules and crystals [46]. The O:H-O as the basic unit of structure and energy to couple the inter- and intramolecular interactions. For the specific situation of water ice compression, computational and spectrometric evidence shows that the O:H−O segmental cooperativity, specific-heat disparity, interionic O—O repulsion, and lone pair polarization dictate the performance of the compressed water ice. Lagrangian transformation of the measured segmental lengths and vibration frequencies resulted in the potential paths of the compressed O:H−O bond [52, 58].

The bonding and electronic dynamics is quite successful. Theoretical analysis revealed the following: i) H−O bond elongation dictates the negatively-sloped $T_C(P)$ boundaries such as the VII–VIII and the Liquid–Quasisolid phase transition for ice regelation; ii) O:H nonbond contraction dominates the positively-sloped $T_C(P)$ profiles such as the Liquid–Vapor phase transition for instant ice formation; and, iii) O:H−O containing angle relaxation governs those of zero-sloped ($T_C$ = constant at the XI–I$_c$ boundary) or the infinitely-sloped ($P_C$ = constant at the (XII, XIII)–X boundary) $T_C(P)$ profiles. Numerical reproduction of the negatively-sloped $T_m(P)$ and $T_C(P)$ for VII–VIII transition results in 3.97 eV of H−O cohesive energy. The O—O repulsion opposes compression, which minimizes water ice compressibility. Polarization enlarges the bandgap and dielectric permittivity of water ice by raising the nonbonding states above the Fermi energy. Progress evidences the efficiency and essentiality of the coupled O:H−O bonding and electronic dynamics in revealing the core physics and chemistry of water ice.

This article is arranged as follows. Section 1 starts with an overview on the anomalies and status regarding the water ice compression derived phase transitioning the phase diagram, dielectrics and bandgap modulation, and the temperature dependence of the compressibility and viscosity. These anomalies inspired our effort made in last decade with an alternative way of thinking and approaching in terms of the coupled bonding and electronic dynamics. Section 2 and 3 describe the general principles for water ice in terms of the coupled O:H−O bond cooperativity and specific heat disparity. Inclusion of the O—O repulsive coupling of the inter- and intramolecular interactions and the introduction of the segmental specific heat has substantiated progress in resolving mysteries of water and ice. Section 4 describes the principle of O:H−O bond and its specific heat response to mechanical compression, which shows how the pressure change the $T_C$ for phase transitions. Section 5 formulates the Grüneisen constant as a function of compressibility, energy density, and bond length. Section 6 shows computational and spectrometric justification of the theoretical predictions of the compression derived bond length and vibration frequency relaxation. Section 7 shows theoretical reproduction of ice regelation and "instant ice" formation by applying mechanical impulse to supercooled water, which gives rise to the O:H−O bond energy of 3.97 eV. Section 8 formulates the O:H length change as a function of pressure along the Liquid-Vapor phase boundary and explains the phase boundaries of constant P and T by showing how the O:H−O bond relaxes crossing these phase boundaries.

Section 9 and 10 demonstrate the joint effect of ionic polarization and mechanical compression on the critical pressures for the room temperature Liquid-VI-VII phase transition, showing that the involvement of interanion repulsion differentiates the effect of solute type from that of solute concentration on the critical $P_C$ trends for the Liquid-VI-VII phase transitions. Section 11 discusses the effect of electrolyte rapid moving impulsion and solute static immersion on the critical temperature of water droplet freezing. Mechanical impulse and ionic solvation effect contrastingly on the O:H−O bond relaxation and the freezing point. Section 13 discusses the compression resolved bandgap expansion and dielectric enhancement of water ice in terms of polarization dominance. Section 14 discusses the temperature dependence of the compressibility, viscosity, and diffusivity of liquid water in terms of the competition between O—O repulsion and thermal depolarization. The last section summarizes the progress and provides a perspective in future endeavor. In each section, a sentence highlighting the key point at the beginning for reader's convenience. Consistency between theory predictions and observations evidences the essentiality and the impact of the coupled O:H–O bond premise and its derivatives to the understanding of water ice compression.

## 2. Principle 1: O:H−O bond cooperativity versus crystallinity

*− O—O repulsion entitles O:H−O a coupled oscillator pair and water a simple molecular crystal*

As the basic functional and interaction elements of molecular crystals, the equally numbered electron lone pairs ":" and dangling protons ($H^+$ simplified as H) entitle water the simplest, ordered, tetrahedrally-coordinated, uniform yet fluctuating structure among known molecular crystals. Represented using a $H_2O:4H_2O$ tetrahedral motif with one $H_2O$ molecular located in the center and four located on the apical sites, water reserves its unique O:H−O configuration and orientation over broad ranges of pressure and temperature, from 5 to 2000 °K and from $10^{-11}$ to $10^{12}$ Pa [19, 59]. $H_2O$ molecular rotation is subject to restriction. Rotating an $H_2O$ molecule around its $C_{3v}$-axis by 120° produces the repulsive H↔H anti-HB and the O:⇔:O super-HB with its nearest neighbors [21], which disorders its two-dimensional hexagonal latticed ice in a long-range order [60]. Dissociating a H−O bond of 5.1 eV energy in the vapor phase requires a laser radiation of 121.6 nm wavelength [61], which constrains the $H^+$ from freely hopping or transitional tunneling from one asymmetrical site to the other of its adjacent oxygen atom.

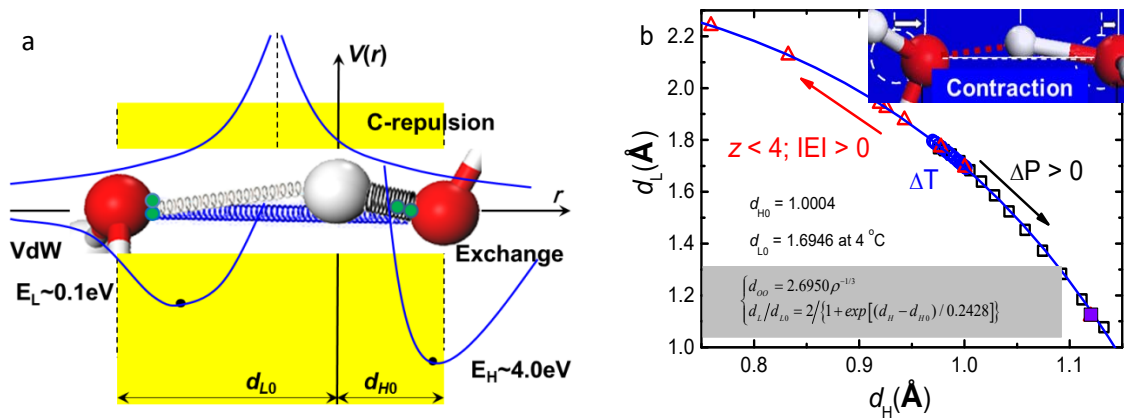

**Fig 1│Potentials of the coupled O:H−O oscillator pair and its segmental length cooperativity** [38, 62]. (a) The O—O repulsive force (C-repulsion) couples the intermolecular O:H nonbond van der Waals (vdW, left-hand side) interaction and the intramolecular H−O polar-covalent bond (Exchange, right-hand side) potential. The rule of (b) O:H−O length cooperative relaxability by multifield perturbation including mechanical compression or tension (P), heating or cooling (T), electrostatic field polarization (E) and molecular undercoordination ($z < 4$). Inset b shows the O:H-O length relaxation by compression and formulation of segmental cooperativity (reprinted with permission from [38, 62])

In placing of the molecular motion scheme, the O:H−O bond, as the basic energetic, functional, and structural unit, features the performance of electrons, bonding, and molecules in the energetic-spatial-temporal domains. Fig 1 illustrates the three-body O:H−O bond potentials and the segmental length cooperativity activated by perturbation. The O:H−O couples the intermolecular O:H nonbond (van der Waals force) and the intramolecular H−O polar-covalent bond (Morse potential) through Coulumb relpulsion between electron pairs of adjacent $O^{2-}$ (C-repulsion). The strong C-repulsion endows the O:H−O an asymmetrical, short-range, and coupled oscillator pair. The choose of the $H^+$ coordination origin and the introduction of the O—O coupling distinct the O:H−O bond from the conventionally referred hydrogen bond, the O:H or the H−O bond alone.

The O:H−O bond relaxes its segmental lengths and energies and its electronic dynamics cooperatively when it is subjecting to a physical perturbation [52]. The inset in Fig 1b formulates the density dependence of molecular O—O separation ($d_{OO}$) and the O:H-O segmental length $d_H \sim d_L$ correlation (subscript x = L for the O:H and H for the H−O) [23]. Under a physical perturbation, the O:H−O changes its length by dislocating both $O^{2-}$ in different amounts $\Delta d_x$ in the same direction. The $H^+$ is the coordination origin of the O:H−O. bond. With the measured molecular separation of $d_{OO} = 2.965$ Å [63], one can readily derive $d_H = 0.889$ Å, $d_L = 2.076$ Å, and $\rho = 0.75$ g/cm$^3$, for the skin of liquid water skin compared with $\rho = 0.92$ g/cm$^3$ at the critical temperature for homogeneous ice nucleation, $T_N = 258$ °K [64].

Mechanical compression shortens the O:H and stiffens its phonons, $\Delta d_L < 0$ and $\Delta \omega_L > 0$, but the H−O responds to compression contrastingly, $\Delta d_H > 0$ and $\Delta \omega_H < 0$, being in contrast to molecular undercoordination [65] or electrostatic polarization [51, 64]. O:H−O only relaxes its segmental length and cohesive energy without alternating the nature or its orientation, unless excessive protons or lone pairs are introduced by solvation or under extreme conditions [66]. Table 1 summarizes the perturbation resolved O:H−O segmental length and vibration frequency relaxation.

The O—O distance determines the coefficient of mass density and polarization determines the surface stress. H−O bond contraction absorbs energy and its length inversion emits energy. Molecular thermal vibration or even evaporation consumes the O:H energy that is caped at ~0.2 eV. As the characteristics of the O:H–O bond, the segmental length, energy, and the stretching vibration frequency ($d_x$, $E_x$, $\omega_x$) of the O:H are about (1.7 Å, 0.2 eV, 200 cm$^{-1}$) and the H−O are about (1.0 Å, 4.0 eV, 3200 cm$^{-1}$) at 277 °K of highest mass density, $\rho_M$ = 1.0 g/cm$^3$ [67]. In contrast, the ∠O:H–O angles, the O—O repulsion, and the segmental lengths are subject to fluctuation that does not vary the statistic values of the bonding characteristics.

Table 1. Coefficients for the O:H–O segmental cooperative relaxation under perturbation in lengths ($d_{H0}$ = 1.0004, $d_{L0}$ = 1.6946 Å), vibration frequencies ($\omega_{L0}$ = 200, $\omega_{H0}$ = 3200 cm$^{-1}$), critical temperatures ($T_{HL}$ = 273, $T_{CL0}$ = 258 K and $T_{LV}$ = 373 K), mass density ($\rho_{M0}$ = 1 g/cm$^3$), and surface stress ($\gamma_0$ = 73.5 J/m$^2$). Einstein's relation $\Theta_{Dx} \propto \omega_x$ applies with respect to $\Theta_{DH0}$ = 3200 K, $\Theta_{DL0}$ = 192 K. Surface stress features the polarization. $T_{CH}$ and $T_{CL}$ are H–O and O:H dictated critical temperatures for phase transitions.

| Bonding | H–O | | | O:H | | | O:H–O | | Ref |
|---|---|---|---|---|---|---|---|---|---|
| q ($\Delta q > 0$)* | $(d_H)_q$ | $(\omega_H)_q$ | $(T_{CH})_q$ | $(d_L)_q$ | $(\omega_L)_q$ | $(T_{CL})_q$ | $(\rho)_q$ | $(\gamma)_q$ | |
| Pressure (P) | > 0 | < 0 | < 0 | < 0 | > 0 | > 0 | > 0 | - | [52] |
| Electric field ($\varepsilon \neq 0$) | | | | | | | | | [68] |
| Molecular CN ($\Delta z < 0$) | < 0 | > 0 | > 0 | > 0 | < 0 | < 0 | < 0 | > 0 | [64] |
| Temperature (T/K) | atmospheric pressure (100 kPa) | | | | | | | | |
| Liquid (277, 377) | < 0 | > 0 | - | > 0 | < 0 | - | > 0 | < 0 | [53] |
| I$_c$ + I$_h$ (100, 258) | | | | | | | | | |
| QS (258, 277) | > 0 | < 0 | - | < 0 | > 0 | - | < 0 | | |
| XI (0, 100) | $\cong 0$ | | | | | | | | |

* $(d_H)_q$ represents for $\dfrac{dd_H}{d_{H0}dq}$ and all the rest the same meaning.

3. Principle 2: *O:H−O* segmental specific-heat: quasisolidity

*− O:H−O segmental energy and vibration frequency define its specific heat and phase boundaries*

In order to describe the thermal response of the O:H−O bond, it is necessary to introduce the segmental specific-heat of Debye approximation, $\eta_x(T/\Theta_{Dx})$ [53]. The specific-heat is the energy cost to raising one °K temperature of the segment. The O:H−O segmental energy and its phonon stiffness uniquely define the $\eta_x(T/\Theta_{Dx})$ curve. The $\eta_x(T/\Theta_{Dx})$ thermal integration from 0 °K to the temperature of its thermal rupture, $T_{Vx}$, is the right cohesive energy $E_x$ of the segment. The Debye temperature $\Theta_{Dx}$ depends directly on the phonon frequency $\omega_x$, following Einstein's relation, $\Delta\Theta_{Dx} \propto \Delta\omega_x$. Hence, any perturbation could mediate the thermodynamics of water ice through the $\eta_x(T/\Theta_{Dx})$ curves by relaxing the $\omega_x$ and $E_x$.

As illustrated in Fig 2a, the segmental $\eta_x(T/\Theta_{Dx})$ of a higher $\Theta_{Dx}$ value approaches to its saturation at a higher temperature. With the given $(\omega_x, E_x, \Theta_{Dx}) = \sim(3200\ \text{cm}^{-1}, 4.0\ \text{eV}, 3200\ °\text{K})$ for the H–O and $\sim(200\ \text{cm}^{-1}, 0.2\ \text{eV}, 192\ °\text{K})$ for the O:H of the bulk water, one can find the relation of $\Theta_{DH}/\Theta_{DL} = 3200/192$. Fig 2a shows the paired $\eta_x(T/\Theta_{Dx})$ curves for the standard reference (broken lines) and for water ice under compression that lowers the $T_m$ and raises the $T_N$ and the $T_V$ through O:H stiffening and H-O softening.

The interplay of the $\eta_x(T/\Theta_{Dx})$ curves results in critical temperatures of the known thermal phase boundaries of water and ice under atmospheric pressure [64], as shown in Fig 2b. Comparing the $\rho(T)$ and the $\eta_x(T/\Theta_{Dx})$ profiles one can find that the segment having a higher $\eta_x$ value undergoes negative thermal expansion, but the other one responds to heating in a regular way. The thermal relaxation of the two segments proceeds in a "master-slave" fashion, leading to the oscillating density anomalies of water ice crossing the phases of Vapor, Liquid, ice-I and XI over the full temperature range.

Most strikingly, the interplay of the $\eta_x(T/\Theta_{Dx})$ curves created an unseen quasisolid phase (QS) of cooling expansion, between Liquid and Ice $I_h$. The QS boundaries ($\eta_L/\eta_H \equiv 1$) match to the extreme densities at temperatures close to the $T_m$ (277 °K, 1.0 g/cm³) and the $T_N$ (258 °K; 0.92 g/cm³) for a large volume. Reducing bulk water into 1.4 nm sized droplet can transits $T_N$ from 258 to 205 °K [53]. It is therefore recommended that the $T_m$ = 277 °K instead of 273 °K at which no signature of density transition is observable.

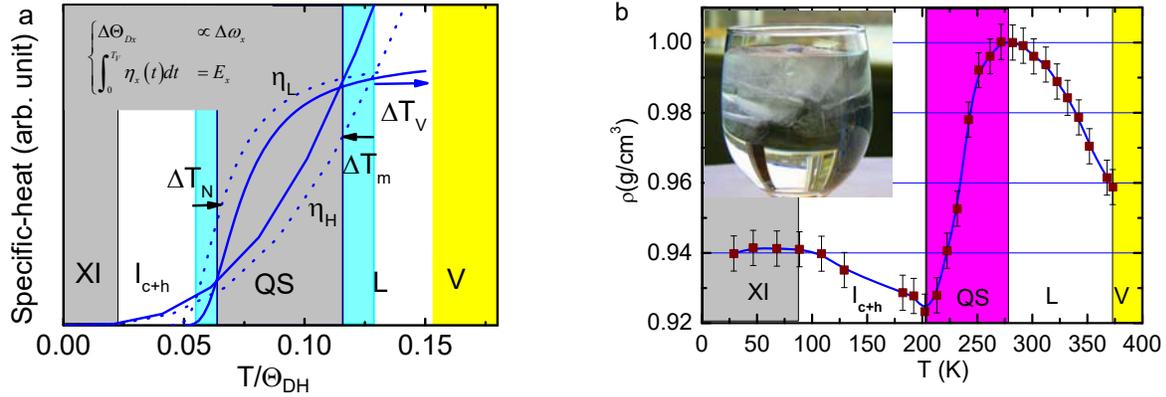

**Fig 2 | O:H−O bond specific-heat disparity versus oscillation of water ice mass density** [53]. (a) Region between the crossing points defines the quasisolid phase (QS) between Liquid and Ice I. The specific heat ratio $\eta_H/\eta_H$ defines the (b) density change over the temperature range and the QS boundaries match the density extremes and the $T_N$ and $T_m$. The $T_V$ is the upper limit of the specific-heat integration. Compression disperses the QS boundary inwardly from the standard reference (broken lines) by O:H contraction and H−O elongation (solid curves), which shift the critical temperatures for melting $T_m$, ice formation $T_N$ and vaporization $T_V$, in manners indicated in (a). (reprinted with permission from [53, 69])

The $\eta_L/\eta_H$ ratio determines the density $d\rho/dT$ slope change. In the QS phase ($\eta_L > \eta_H$), cooling contraction of the $d_H$ masters the O:H−O relaxation and mass density variation. The $d_H$ contracts less than the $d_L$ expands, and thus the density reduces gradually to a minimum at the $T_N$ for ice formation. The observation clarifies the origin of ice floating though the ∠O:H−O angle expands from 160 to 167 ° according to quantum computations [53].

In the Vapor phase ($\eta_L \cong 0$), the gaseous molecule can be taken as the structural unit because the O:H interaction is negligibly weak. The H−O bond is shortest of 0.09 nm and strongest of 5.1 eV energy. In the Liquid and the $I_{c+h}$ phases ($\eta_L/\eta_H < 1$), the $d_L$ masters the relaxation. The $d_L$ undergoes cooling contraction and the $d_H$ expansion, so the mass density increases in the Ice I and the Liquid phase at different rates. Although the Liquid and the Ice $I_{c+h}$ phases undergo the regular thermal expansion, the mechanism is entirely different from and much more complicated than one could imagine for regular substance. The $d_H$ thermal contraction absorbs energy and its reverse does oppositely, which forms the ground for the Mpemba effect – warm water cools more quickly [70]. In the lower temperature XI phase, the Debye form of the $\eta_x(T/\Theta_{Dx})$ closes zero, $\eta_L \cong \eta_H \cong 0$, neither the O:H−O segmental lengths nor their energies are sensitive to temperature change, giving rise to $\Delta\omega_x \cong 0$ [71, 72]. The ∠O:H−O angle undergoes cooling expansion from 167 to 175 °, which only slightly lowers the mass density of the ice XI phase, as reported in Ref [69].

## 4  Principle 3: O:H−O length symmetrization and its coupled potentials

*− Lagrangian solution transforms the segmental length and stiffness to the potential paths*

Developing the hydrogen-bond interaction potential is of vital importance to understanding the biological systems and to engineering the X:H–Y bond in general (X, Y = N, O, F could generate lone pairs in reaction) [73-75]. Teixeira [50] proposed in 1998 a symmetrical double-well potential that accommodates the $H^+$ proton "frustrating" between adjacent oxygen [46]. Wernet et al [76] and Kumagai [77] proposed an asymmetrical "double-well" potential for the transition tunneling of the proton in the process of $2H_2O \rightarrow OH_3^+ : OH^-$ transition. When the $O^{2-}$ ions are pushed closer, the symmetrical double-well potential transits into single well located at a site midway between the adjacent $O^{2-}$ ions. However, neutron and X-ray diffraction, photoelectron emission, phonon spectroscopy could hardly validify the proposal of asymmetric potentials. In 2013, Huang at al [58] developed the asymmetrical and short-range potential for the coupled O:H−O oscillator pair, in which the $H^+$ is taken as the coordination origin, as shown in Fig 1a.

MD computations [52] resolved the $d_x(P)$ profiles by reproducing the V-P equation-of-state of 80 °K ice [52], see eq (5) and Fig 3 a. The $d_x(P)$ profiles meet at 59 GPa and reproduce the identical $d_H$ and $d_L$ length of 0.11 nm [13, 16]. These $d_x(P)$ profiles yielded the $d_{OO}$ - ρ and $d_H$ - $d_L$ correlations by taking the tetrahedron geometry [62], as formulated in Fig 1 b inset.

Lagrangian-Laplacian resolution to the motion mechanics of the coupled O:H−O oscillator pair converted the known $(d_x, \omega_x)$, see eq (5), into the force constant and cohesive energy, $(k_x, E_x)$, leading to the potential paths of the compressed O:H−O bond. Fig 3 b plots the potential paths of the O:H−O relaxing under compression. The potentials have two key characteristics compared with available models. One is the location of the H$^+$ coordinate origin and the other is the involvement of the O—O repulsion force. The small blue circles are the initial equilibrium of O$^{2-}$ ions without the Coulomb repulsion or compression being involved. The open red circles are the quasi-equilibrium states at each value of pressure. The cooperativity of the O:H and the H−O potential paths proved the effect of both mechanical compression and the Coulomb repulsion between adjacent oxygen ions on the segmental length and energy.

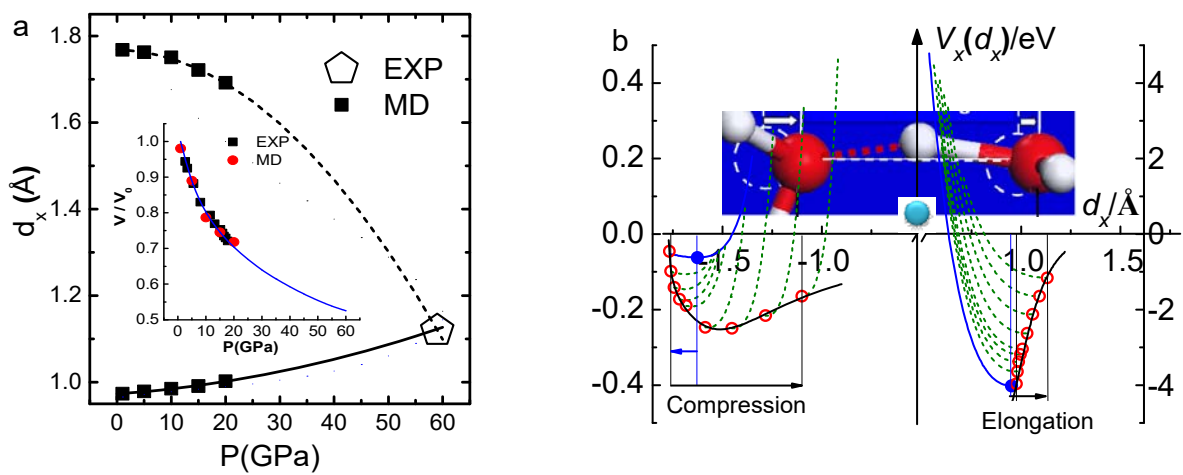

**Fig 3 | Compression symmetrized segmental length and the coupled O:H−O potential paths** [52, 58]. (a) MD derived $d_x$–$P$ curves from the V-P profile of 80 °K ice (inset) [23], which meet at the ice VIII-X phase boundary of identical O:H (upper part) and H-O length (lower part) of 1.10 Å under 59 GPa compression [13, 16]. The upper/lower part corresponds to the $d_L/d_H$ segment. (b) O:H−O bond potential path functions $V_x(r)$ of the compressed ice (from left to right with $P$ increasing from 0 to 60 GPa). Compression dislocates both O$^{2-}$ ions to the right-hand side because of the involvement of O—O repulsive coupling. Note the scale difference in vertical axis for the O:H and H-O energy. (reprinted with permission from [52, 58])

The O—O repulsion dislocates both O$^{2-}$ ions toward the right-hand side direction under compression, rather than they move closer, as one initially thought. The vertical energy scales are one order difference between the two axis. Depending less on the probing pressure, the intrinsic $V_x(r)$ potential paths feature the right asymmetrical, coup-led, and short-range interactions for the coupled O:H−O oscillator pair, which can be extended to organic molecular crystals having lone-pair interactions. Contrastingly, the O:H−O bond potential paths for undercoordinated molecules undergo O:H elongation and H−O contraction [78].

## 5   Analytical forms of bond relaxation

*− O:H−O segmental relaxation discriminates the critical energy for a specific phase transition*

Physical perturbations such as compression ($P$), atomic undercoordination ($z$), temperature ($T$), and electrostatic polarization ($\varepsilon$) relax the length, stiffness (vibration frequency), and energy of the representative bond of a specimen. Bond relaxation mediates the electronic energetics by entrapment and polarization, which determines the measurable properties of a substance under perturbation. Any perturbation transmits the potential from the equilibrium $U(d_0)$ to a new equilibrium $U(d_0)(1 + \Delta)$ by bond length and energy relaxation [79],

$$\begin{cases} d(P,T) = d_0 \left[ 1 + \dfrac{\left(1 + \int_{T_0}^{T} \alpha(t)dt\right)\left(1 - \int_{P_0}^{P} \beta(p)dp\right) \cdots}{d_0} \right] \\ E(P,T) = E_0 \left[ 1 - \dfrac{\int_{T_0}^{T} C_v(t)dt/z - \int_{V_0}^{V} p(v)dv}{E_0} \right] \end{cases}$$

(2)

The $\alpha(t)$ is the thermal expansibility. The $\eta(t) = C_v(t/\theta_D)/z$ is the reduced specific heat of the representative bond and z the is the atomic coordination number per atom (the $C_v(t/\theta_D)$ has been divided by the Boltzmann constant 3R for the bulk specimen). The $\beta = -\partial v/(v\partial p)$ is the compressibility that is proportional to the inverse of elastic bulk modulus. The p(V) is the equation of states. The Grüneisen parameter is often used to quantify the response of the phonon frequency to the stimulus from measurements. According to the expression of the relation, $\omega^2 \propto E/d^2$, one can express the Grüneisen parameter for a representative bond as a function of its bond length, binding energy density and atomic cohesive energy, shown as follows:

$$\gamma_q = \frac{d\omega}{\omega_0 dq} = \frac{dE}{2E_b dq} - \frac{dd}{d_b dq}$$

$$\Rightarrow \begin{cases} \gamma_T = \dfrac{d\omega}{\omega_0 dt} = -\alpha(t)\left[\dfrac{d_b F(d_b + \delta)}{2E_{coh}} + 1\right] \\ \gamma_P = \dfrac{d\omega}{\omega_0 dp} = \beta(p)\left[\dfrac{P}{2E_{den}} - 1\right] \end{cases}$$

where

$$\begin{cases} \dfrac{-\beta P}{E_{den}} = \dfrac{Pdv}{E} = \dfrac{VPdv}{EVdp} = \dfrac{VP}{E}\left(-\dfrac{dv}{Vdp}\right) \\ \alpha(t) \cong \dfrac{dL}{L_0 dt} = \dfrac{1}{L_0}\left(\dfrac{\partial L}{\partial U}\right)\dfrac{dU}{dt} \cong \dfrac{-C_v(t)}{zL_0 F(r+\delta)} \end{cases}$$

(3)

$E_{coh} = zE$ is the atomic cohesive energy. Where $dU/dL = -F(d+\delta)$ is the gradient of the interatomic potential $U(d+\delta)$ upon thermal relaxation from the equilibrium d to $d + \delta \sim d(1 + 3\%)$ at melting [80]. The $dU/dt$ is the specific heat $C_v$. Therefore, the Grüneisen parameters are connect to the intrinsic thermal expansivity (specific heat $C_v$) $\alpha$, compressibility $\beta$, atomic cohesive energy $E_{coh}$, bond length d and bonding energy density $E_{den}$. P is a constant with a sign for its direction. Given proper conditions one can find of these quantities from the Grüneisen parameters derived from spectral measurements. Besides the constant -1, the sign of the $\gamma_P$ is always positive because of the product $\beta P > 0$ for regular substance; the $F(d_b + \delta)$ determines the sign of the $\gamma_T$ because of the positive specific heat, $\eta_v > 0$. The $F(d_b + \delta)$ holds the same sign of the bond length increment $\delta$. Therefore, thermal contraction gives rise to the positive $\gamma_T$.

The critical temperature $T_C$ for a certain phase transition is proportional to the segmental cohesive energy $E_C$. We thus have the expression for the pressure dependence of the $T_C$,

$$\frac{dT_C(p)}{T_{C0}dp} = \frac{dE_C(p)}{E_{C0}dp} = -\frac{pdv}{E_{C0}dp} = -\frac{Vpdv}{E_{C0}Vdp} = \frac{\beta p}{E_{denC0}}$$

(4)

For the 'normal' substance, the critical $T_C$ for phase transition is proportional to the atomic cohesive energy, $T_C \propto zE_z$, where z is the number of bonds to a z-coordinated atom and $E_z$ is the bond energy [81].

However, for the O:H−O bonded systems, the $T_{Cx}$ is proportional to the single-bond energy $E_x$ because surrounding lone pairs isolate the H$_2$O molecule, and $z/z_b = 1$. Furthermore, The O:H and H−O respond to perturbation contrastingly in length and energy, which discriminate the $T_{Cx}$ for a phase transition from another in the phase diagram, see Fig 4 b. Fig 4 a shows the O:H−O segmental energy exchange upon compression with noted $P_{C1}$ and $P_{C2}$ for consecutive phase transitions. Compression raises the O:H energy while reduces the H−O energy, agreeing with quantum computations [25].

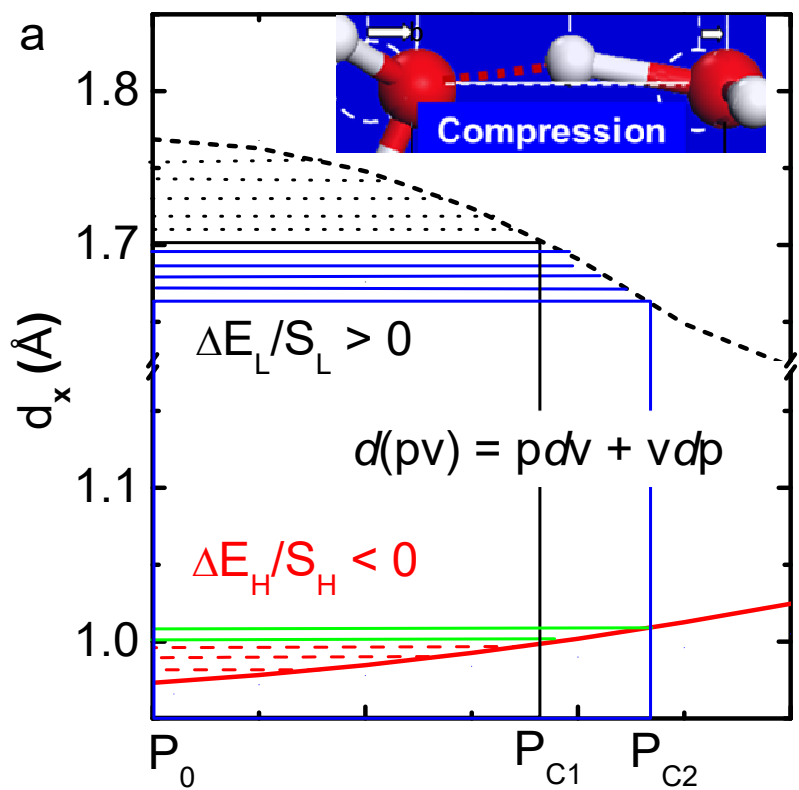

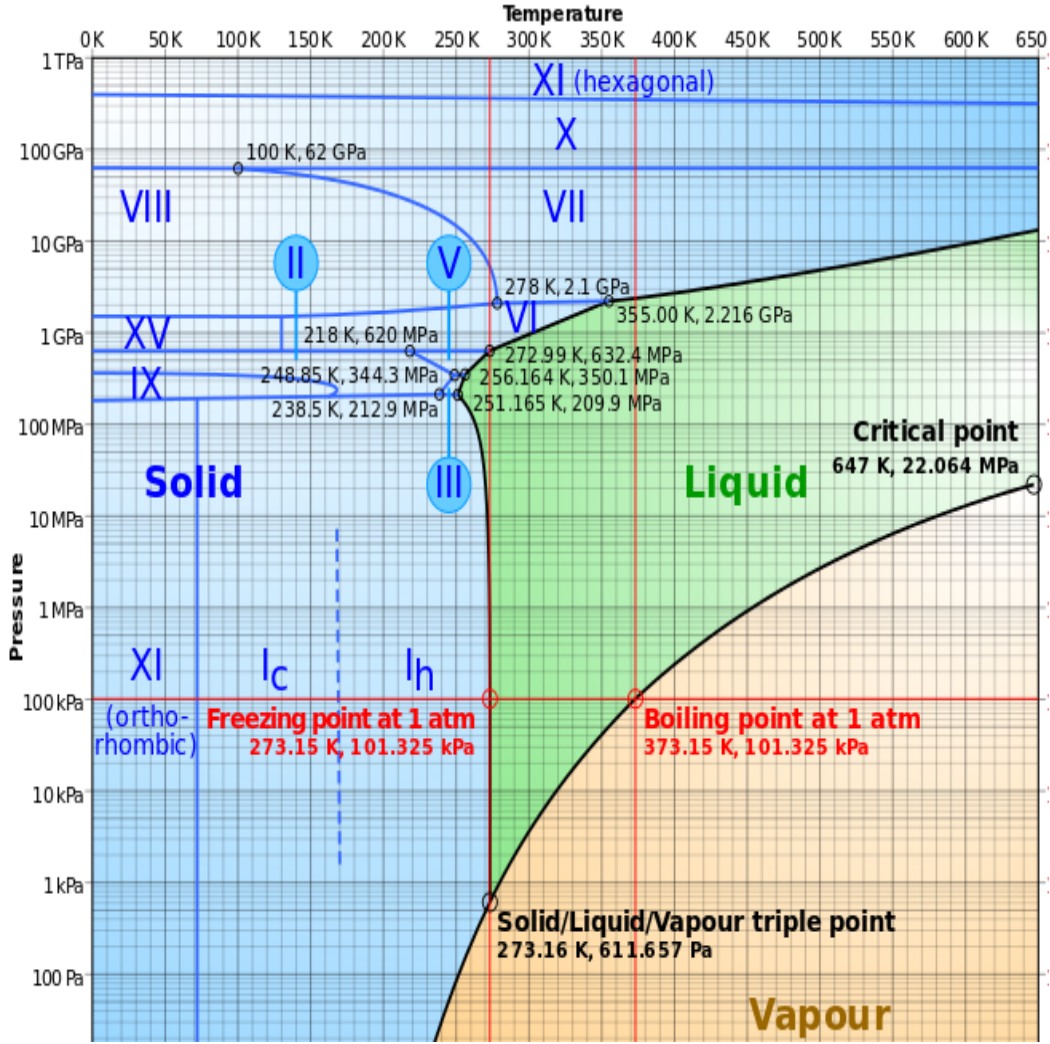

**Fig 4 | O:H−O segmental cooperative $d_x$ and $E_x$ relaxation discriminates the phase boundaries** [68, 82]. (a) Shaded area of the upper branch corresponds to the O:H energy gain and the lower branch to H−O energy loss under compression, which discriminate b) phase boundaries according to their $dT/dP$ slopes. The $S_x$ is the cross-sectional area of the $d_x$ segment. Upper/lower part corresponds to the $d_L/d_H$ segment. (reprinted with permission from [68, 82])

6   O:H−O phonon frequency relaxation and phase transition

*− Compression symmetrizes O:H−O except for phase boundaries or extremely high pressure*

Fig 5 shows the Raman phonon spectra of the 298 °K deionized water under pressure. The spectra show the sequential Liquid-VI-VII phase transition. Spectral features centered at 75 cm$^{-1}$ and 3450 cm$^{-1}$ correspond to the undercoordinated skin O:H−O bonds and those around 300 and 3200 cm$^{-1}$ from the bulk water. The $\omega_H$ at 3150 cm$^{-1}$ comes from bulk ice. The phonons of the weaker O:H nonbond ($\omega_L$ < 500 cm$^{-1}$) undergo blueshift and the phonons of the stronger H−O bond ($\omega_H$ > 3000 cm$^{-1}$) undergo redshift except for phase boundaries. The 3450 cm$^{-1}$ skin signature persists throughout the course of detection, which evidences that the shorter skin bond is insensitive to compressive perturbation. Inset (b) shows an optical image of the room-temperature ice formed at 2.17 GPa.

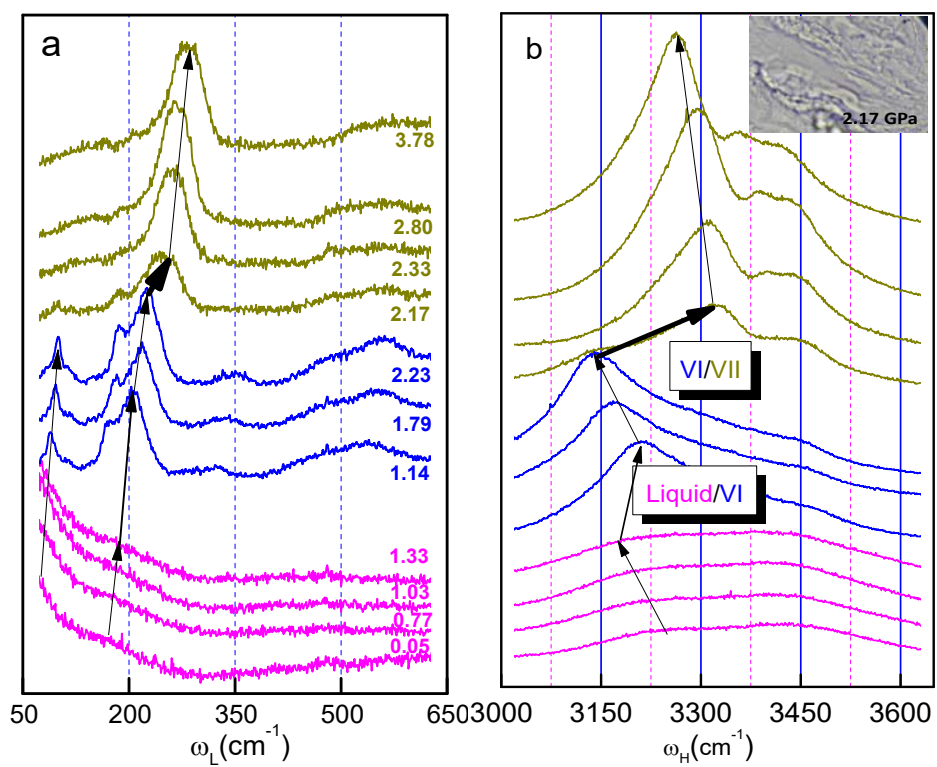

**Fig 5 | Raman spectra of the 298 °K water compression icing transition** [68]. Vertical axis is the spectral intensity (arbitrary unit) offset for clarity. Transition from Liquid to Ice-VI at 1.33-1.14 GPa and from VI to VII at 2.23-2.17 GPa show consistently O:H stiffening and H−O stiffening except for the phase boundaries. Inset (b) shows the optical image of ice at 2.17 GPa. The abrupt pressure drop at phase boundaries and the simultaneous $\omega_L$ and $\omega_H$ blueshift at phase boundaries shows the repulsive O—O degeneration upon structure evolution. (reprinted with permission from [68])

Phase transition from Liquid to Ice−VI happens at 1.33 ~ 1.14 GPa and then to ice-VII at 2.23 ~ 2.17 GPa.

The $\omega_L$ blueshift and $\omega_H$ redshift happen simultaneously throughout all phases, evidencing that O:H contraction and H−O elongation pertain to the compressed water ice. The abrupt pressure drop and the simultaneous $\omega_L$ and $\omega_H$ blueshift, particularly at the VI−VII boundary, shows that repulsive O—O degenerates and then recovers upon the VII phase is formed.

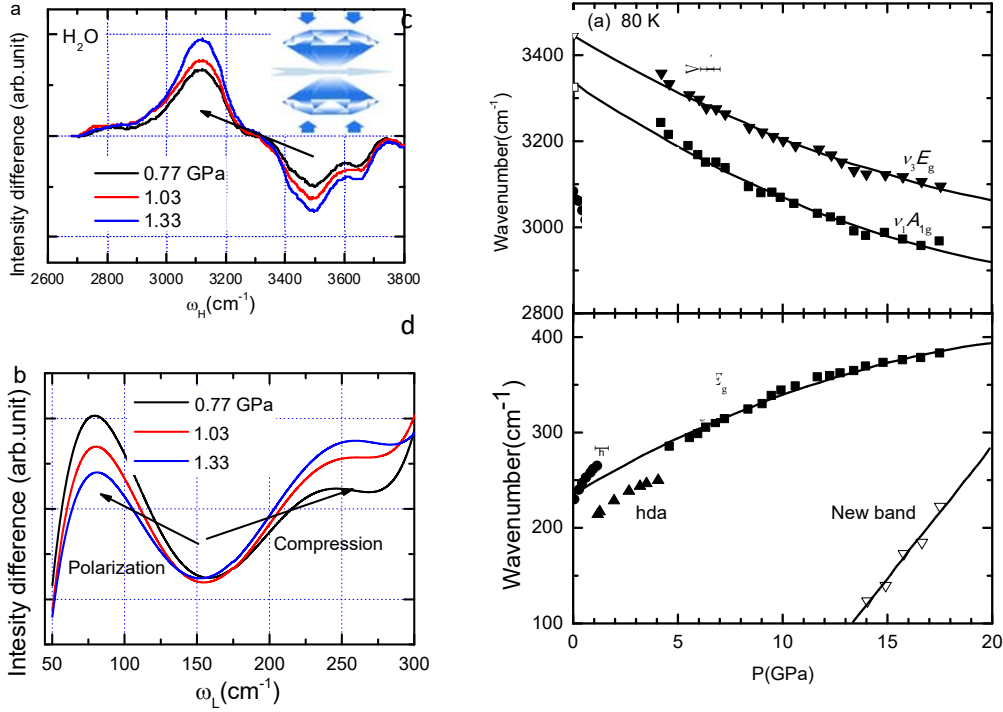

**Fig 6 | Compression-resolved $\omega_x$ shift for ambient water** [68] **and 80 K ice** [23]. Compression softens (a, c) the $\omega_H$ and stiffens (b, d) the $\omega_L$ monotonically for (a, b) the ambient water (derived from Fig 5) and (c, d) 80 °K ice. (reprinted with permission from [23, 68])

The differential phonon spectra (DPS) of the 298 °K water in Fig 6 a and b, show that compression transits the $\omega_H$ from 3500 cm$^{-1}$ to 3100 cm$^{-1}$ and derives two $\omega_L$ satellites of polarization at 75 cm$^{-1}$ and the compressed mode at 250 cm$^{-1}$ and above. The DPS is obtained by subtracting the reference spectrum collected at the ambient pressure from those compressed upon all the spectral characteristic peaks being area normalized, which distills the effect of a perturbation on the $\omega_x$ shift [55]. Fig 6 c and d show the phonon frequency shift for the compressed 80 °K ice crossing the I$_c$, IX, II, XV, and VIII phases [23]. The bulk $\omega_H$ at 3310 and skin $\omega_H$ at 3450 cm$^{-1}$ undergo redshift and the $\omega_L$ shits to higher frequencies throughout the course of detection. Other minus signatures are out of concern in the present analysis. The slope of the $\omega_H$-P curve shows the negative $\gamma_{PH}$ that is more linear dependence than the positive $\gamma_{PL}$ on pressure because of the softer O:H is more sensitive to the O—O repulsion and compression. The following presents the $\omega_x$ and $d_x$ fitting to measurements with derivative of the negative $d_H$ compressibility and Grüneisen parameter [52].

$$\begin{pmatrix} d_H/0.9754 \\ d_L/1.7687 \\ \omega_H/3326.14 \\ \omega_L/237.422 \end{pmatrix} = \begin{pmatrix} 1 & 9.510\times 10^{-2} & 0.2893 \\ 1 & -3.477\times 10^{-2} & -1.0280 \\ 1 & -0.905 & 1.438 \\ 1 & 5.288 & -9.672 \end{pmatrix} \begin{pmatrix} P^0 \\ 10^{-2}P^1 \\ 10^{-4}P^2 \end{pmatrix} \Rightarrow \begin{pmatrix} \beta_{PH}<0 \\ \beta_{PL}>0 \\ \gamma_{PH}<0 \\ \gamma_{PL}>0 \end{pmatrix}.$$

(5)

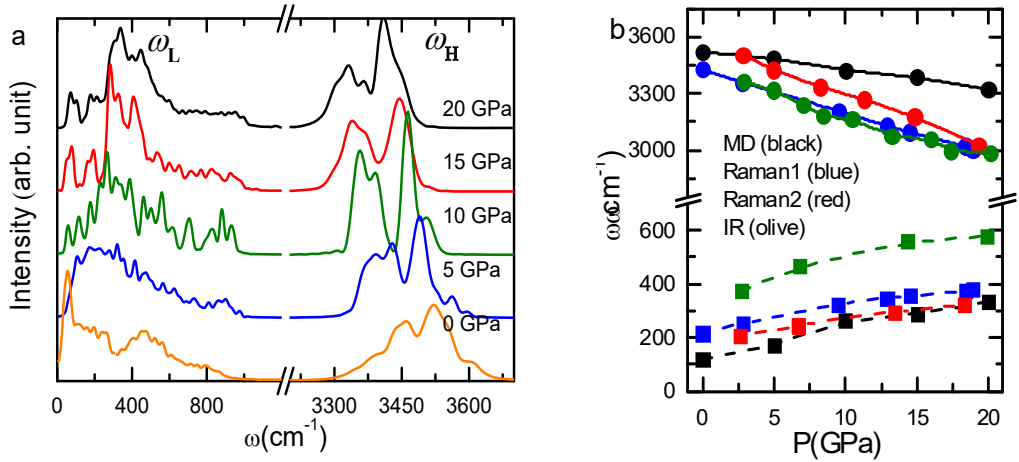

**Fig 7 | O:H−O segmental phonon frequency shift** [52]. (a) MD-derived $\omega$ spectra and agree well to the (b) Raman and IR spectroscopic trends, which evidences further the effect of compression on the H−O mode softening and the O:H mode stiffening for the 80 K ice−VIII phase [11, 12, 23]. (reprinted with permission from [52])

Fig 7a displays the MD derived phonon spectra of the compressed ice-VIII up to 20 GPa, which agrees with phonon spectroscopic trends measured from 80 K ice (Fig 7b) [11, 12]. Compression shifts the $\omega_H$ from 3520 cm$^{-1}$ to 3320 cm$^{-1}$ and the $\omega_L$ from 120 to 336 cm$^{-1}$, regardless of what the structural phase is. Consistency in the pressure-induced $\omega_x$ cooperative relaxation for both ice [10-12] and water [24] clarifies that compression shortens and stiffens the O:H and lengthens and softens the H−O bond in liquid water and ice.

## 7 Regelation and instant ice formation

*− H−O bond energy dictates the $T_m$ while the O:H energy determines the $T_N$ and $T_V$*

One may categorize the $T_C(P)$ phase boundaries in Fig 4 b, into four groups to reproduce the boundary profiles according to their slopes [22]:

$$\frac{dT_C(P)}{dP} = \begin{cases} \cong \delta(P_C) & [\beta_{PH}/E_{den,H0} + \beta_{PL}/E_{den,L0} \cong 0] & (\text{Parellel to } T\text{-axis}) \\ \cong 0 & (\beta_{Px} \cong 0) & (\text{Parellel to } P\text{-axis}) \\ > 0 & (\beta_{PL} > 0) & (Liq - Vapor,...) \\ < 0 & (\beta_{PH} < 0) & (Liq - QS, VII - VIII,...) \end{cases}$$

(6)

Where δ($P_C$) is the Kronig singularity function. If $P = P_C$, δ($P_C$) = 1, otherwise, δ($P_C$) = 0.

From dimensional point of view, the $T_C$ corresponds to a certain form of energy. Phase transition needs to loosen these bonds to a certain extent. For instance, one needs to break all bonds of an atom to its neighbors for evaporation. However, for water ice, a $H_2O$ molecule is isolated by its surrounding lone pairs and other neighbors contribute to the cohesive energy insignificantly. So, the $T_C$ is proportional to the $E_L$ or the $E_H$ of the molecule for a specific transition. Eqs (4) and (6) describe the $T_C(P)$ in terms of the O:H−O segmental volume under compression, $dT_C \propto -pdv_x$. The following will discuss how the $E_x$ dictates the categorized phase transitions, according to the slope of the phase boundary.

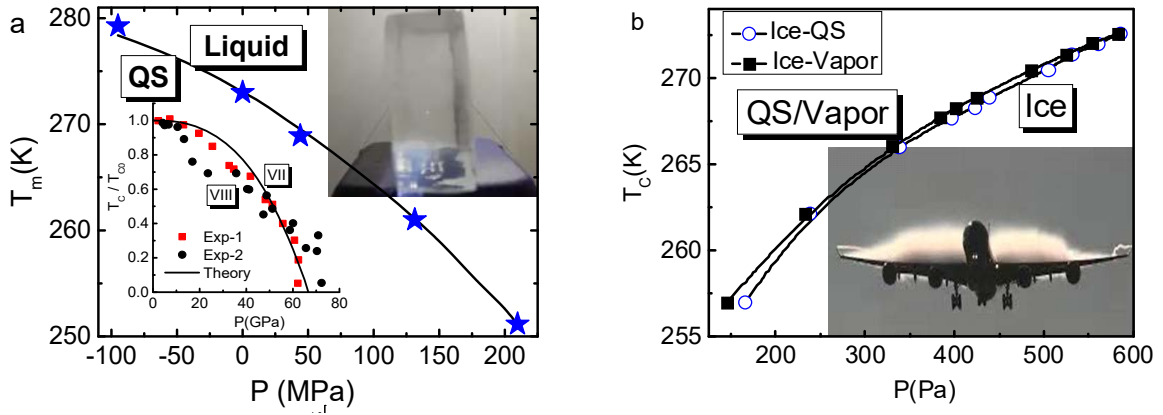

**Fig 8 | Theoretical reproduction of the QS–Liquid and VII–VIII, Liquid–Vapor, and QS–Ice phase boundaries** [52, 83]. Compression disperses the QS boundary inwardly to (a) lower the $T_m$ for Regelation (inset a) [5] and (b) raises the $T_V$ for (QS, Ice)/Vapor transition and the $T_N$ for "instant ice" formation [9]. Inset b shows Vapor–Ice transformation on the upper side of the flying aircraft wings under low-pressure and low-temperature.(reprinted with permission from [52, 83])

Fig 8 shows reproduction of phase boundaries with finite $T_C(P)$ slopes using eq (4). Using Michelson interferometry and laser confocal microscopy, Chen et al [9] confirmed that the $T_C$ for the Ice-QS ($T_N$) and Ice-Vapor ($T_V$) transition increases with pressure. The negative slopes of the QS–Liquid ($T_m$) and the VII–VIII boundaries indicate a negative compressibility, and positive slopes for the (Liquid, Ice)–Vapor ($T_V$) indicates a positive compressibility. The Ice–QS transition ($T_N$) follows the same pressure trend of $T_V$. The $T_m(P)$ curve defines the Regelation – compression lowers the $T_m$; the $T_V(P)$ line features vaporization and the $T_N(P)$ the formation of "instant ice". The $T_V$ for evaporation is the upper limit of the O:H specific-heat integration, as the O:H interaction in the vapor phase becomes negligible.

Let us look at the $T_C(P)$ phase boundaries of finite slopes,

$$\frac{\Delta T_C(P)}{T_C(P_0)} = \frac{1}{E_{coh,x0}} \int_{P_0}^{P} \beta_x(p) p \, dp \begin{cases} <0 & (T_m, T_C(VII-VIII); \beta_H < 0) \\ >0 & (T_N, T_V; \beta_L > 0) \end{cases},$$

with the known $dd_H/dp = -\beta_{PH} > 0$ and $dd_L/dp = -\beta_{PL} < 0$, given in eq (5).

(7)

Reproduction of the $T_C(P)$ curves in Fig 8, using the $d_x(P)$ given in eq (5), confirmed that the $E_H$ dictates the negatively-sloped $T_m$ for ice regelation (251 °K at 210 MPa; 283.5 °K at -95 MPa) [8, 84] and the $T_C$ for the VII–VIII phase transition [10-12]. The $E_L$ determines the $T_V$ and $T_N$. Fitting to both the $T_m(P)$ profile and the $T_C$ for the VII–VIII transition yields the $E_H$ value of 3.97 eV by taking the H atom diameter 0.106 nm as that of the H−O bond [52]. With the known bulk $E_L \sim 0.2$ eV and $dd_L/dp < 0$, one can derive the effective cross-section area $S_L$ for the O:H nonbond.

One can thus understand the ice Regelation and "instant ice" formation from the perspective of QS boundary dispersion by mechanical compression. Compression raises the $T_N$ and $T_V$ and lowers the $T_m$ through the $\Delta\Theta_{Dx} \propto \Delta\omega_x$ relation. Compression and polarization split the O:H phonon band into the 75 and 250 cm$^{-1}$ components and shift the H−O vibration frequency from 3500 to 3100 cm$^{-1}$ and below, which shift the $\Theta_{DX}$ and these $T_C$ values accordingly. The $T_m$ depression is associated with $T_N$ and $T_V$ elevation. Indeed, the $T_N$ for ice nucleation and the $T_V$ for ice-vapor transition increase from 258 to 273 °K when the pressure is increased from 150 to 600 Pa [9].

The $T_N$ elevation explains how the "instant ice" is formed by applying a mechanical shock impulse $\Delta P > 0$, as Fig 8b inset shows. One can also slap a bottle of cooled water to raise the $T_N$ by supplying a pressure impulse. The $T_N$ and $T_V$ are much more sensitive to a compressive perturbation than the $T_m$. It is also clear why the T$_V$ becomes lower of water on the high latitude such as mountain tops having a pressure lower than the atmospheric, which should be accompanied with a higher T$_m$ and a lower T$_N$. Fig 8b inset shows vapor transition into ice on the wings of a flying aircraft. Following Bernoulli's equation, the low-pressure $P = P_0 - \rho(gH + 0.5v^2)$, depresses the T$_N$ for ice formation on the top side of the wings where the temperature is some 233 °K or even lower. The ρ is the vapor mass density, H the height, g the gravity and v the velocity. P$_0$ is the atmospheric pressure.

8      Liquid−Vapor transition and constant phase boundaries

*− O:H or H−O relaxation discriminates phase boundaries according to their dT/dP slopes*

We show now that O:H relaxation dictates the $T_V(P)$ and derive the $d_x(P)$ function. The $dd_L/dp < 0$ meets the criterion of $dT_C/dP > 0$ for the Vapor−(Liquid, Ice) phase boundaries. The following expression fits to the Liquid-Vapor transition $T_C(P)$ curve [82], see the phase diagram in Fig 4b, for instance,

$$\frac{\Delta T_C(P)}{225.337} = 0.067757 \times \exp\left(\frac{LnP}{5.10507}\right) = AP^{\frac{1}{B}}; \quad \Delta T_C(P) = 17.52153 P^{0.195884}$$

(8)

Equaling eqs (7) and (8), yields the pressure dependent O:H length,

$$\frac{\Delta d_L(P)}{d_{L0}} = \frac{4.2683 E_{L0} P_0^{-0.80412}}{T_C(P_0) v_{L0}} \left[\left(\frac{P}{P_0}\right)^{-0.80412} - 1\right] \propto \left[\left(\frac{P}{P_0}\right)^{-0.80412} - 1\right]$$

(9)

The O:H relaxation governs the process of vaporization, in which the $d_L$ varies with pressure in a form of $P^{-0.8}$ dependent. The $v_{L0}$ is the volume of the O:H bond at the standard ambient reference.

The following describes phase boundaries that are either parallel or perpendicular to the $T$-axis in the phase diagram,

$$\frac{\Delta T_C(P)}{T_C(P_0)} = \sum_{H,L} \frac{\int_{P_0}^{P} \beta_{Px}(p) p \mathrm{d}p}{E_{den,x0}} = \begin{cases} \delta(P_C), & \left(\frac{\beta_{PL}(p)}{E_{den,L0}} + \frac{\beta_{PH}(p)}{E_{den,H0}} = 0;\ T_C = const\right) \\ 0, & (\alpha_x(t) = 0;\ \eta_x \cong 0;\ P_C = const) \end{cases}$$

(10)

The X – (XI, VII, VIII) transition stands for the first case of constant $T_C$ at higher pressures and the high-$T_C$ is insensitive to heating. The $d_H$ and the $d_L$ are identical when enters the VIII–X phase under 60 GPa pressure [52]. Computations revealed that further compression shortens both the H−O and the O:H length slightly when cross the VIII–X phase boundary, reaching a similar and higher force constant without de-hybridizing the *sp* orbits of oxygen [85]. Mechanical compression compensates the heating O:H elongation and the H−O contraction along the X – (XI, VII, VIII) boundaries [41] with the $\beta_x(p)/E_{den,x0}$ opposite signs, and then both O:H and H-O are compressed.

The fourth situation is the XI – I$_C$ transition at 70 °K. The boundary is insensitive to pressure and perpendicular to the T-axis. The zero $T_C(P)$ slope means the energy independent of the phase boundary, none of the O:H or the H−O undergoes thermal relaxation, $\mathrm{d}d_x/\mathrm{d}T \cong 0$, because the specific-heat $\eta_x \approx 0$ at very low temperatures. Only the ∠O:H—O angle enlargement contributes to the structure relaxation without substantial energy or stretching phonon frequency change [53].

Thus, the O:H-O cooperativity provides hither to bonding dynamics comprehension of the T-P phase diagram over all phases and along all phase boundaries. The negatively sloped boundaries are dominated by H-O relaxation and the positively sloped ones by O:H relaxation. The ones along the P-axis arise from the O:H-O frozen and the ones along the T-axis results from the O:H and H-O compensate relaxation, $\beta_{PL}(p)/E_{den,L0} + \beta_{PH}(p)/E_{den,H0} = 0$. One can formulate how the O:H or the H-O length change with pressure, such as the case of Vapor-Liquid boundary (eq 8).

9	Pure and salt ices compression: O:H−O polarization against compression

*− Compression can hardly lengthen the already-shortened H−O bond further*

Fig 9 shows that the H–O phonon frequency and the O—O distance for the salt solutions have smaller and negative relaxation coefficients (slope), as expected. Fig 9a shows that mechanical compression softens the $\omega_H$ phonon regardless of the presence of LiCl solutes [86]. Recent progress revealed that ions occupy the hollow sites to form each a $(\pm)\cdot 4H_2O:6H_2O$ motif (Fig 9a inset) [87]. The ionic electric field aligns, polarizes, and stretches water dipoles to form a hydration volume. The hydrating H−O bond is shorter and stiffer. Therefore, the higher frequency phonons arise from those H−O bonds inside the hydration volume. The initially shorter and stiffer H–O bonds inside the hydration volume are mechanically and thermally more stable than they are in the ordinary water and ice [16, 17]. So, the $\omega_H$-P (solid) curves of hydrating H-O bonds show the negative and smaller $\gamma_{PH}$ slopes than those outside the hydration volume.

Fig 9b compares the O—O distance (O:H–O length) in ices with and without (Li, Na)Cl solutes [88]. The hydrating O:H–O bond in the $Li^+$ and $Na^+$ hydration volume is always longer than those outside. These observations confirm the contrasting effects of ionic polarization and mechanical compression on the O:H–O segmental length and stiffness cooperative relaxability.

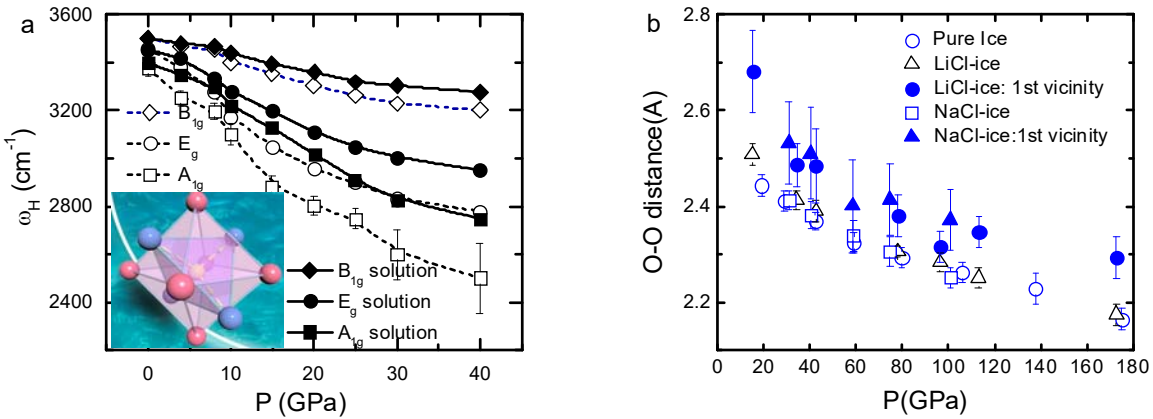

**Fig 9 | Ionic polarization opposes compression in softening the H−O and shortening the O—O** [88]. (a) Compression and ionic hydration resolved H–O phonon relaxation in pure (broken lines) and 0.08 mol concentrated LiCl/H$_2$O (solid lines) ices [86] and (b) the molecular site and pressure resolved O—O distance in pure and (Li, Na)Cl/H$_2$O ices. Inset a shows the ionic (±)·4H$_2$O:6H$_2$O hydration volume [87]. The B$_{1g}$, E$_g$, A$_{1g}$ are vibration modes based on group symmetry. (reprinted with permission from [88])

## 10      Compression icing of room-temperature salt solutions

*− Compression recovers the ionic polarization induced O:H–O relxation for L–VI and VI–VII transition*

### 10.1     Principle: compression and polarization compensation

To compensate the effect of polarization, one needs higher pressure to transit a salt solution from one phase to another occurred to water ice. The O:H–O bond energy $E_{Cx}$ correlates the $P_{Cx}$ and the $T_{Cx}$ for the phase transition (x = L, H) of pure water and salt solution at the same $T_{Cx}$:

$$T_{Cx} \propto E_{Cx}$$
$$= \begin{cases} E_{x0} + v_x \int_{P_0}^{P_{C0}} \beta_x(p) p \, dp & \text{(a, neat } H_2O) \\ E_{x0} + \int_0^E \kappa_x(\varepsilon) \varepsilon \, d\varepsilon + v_x \int_{P_0}^{P_C} \beta_x(p) p \, dp & \text{(b, solution)} \end{cases}$$

(11)

One needs to raise the $P_C$ from $P_0$ to $P_{C0}$ to overcome the effect of electrification when turns pure water to a slat solution for the same phase transition at the same temperature (ΔT$_{Cx}$ = 0). The second term for solution is the O:H–O bond energy storage by ionic polarization with κ being the dielectric permittivity,

$$\Delta E_x = \int_0^E \kappa_x(\varepsilon) \varepsilon \, d\varepsilon = -v_x \int_{P_{C0}}^{P_C} \beta_x(p) p \, dp = \begin{cases} > 0 & (H-O) \\ < 0 & (O:H) \end{cases}$$

(12)

Increasing the $P_{C0}$ to $P_C$ recovers the bond distortion by ionic polarization, and then phase transition takes place. If one wishes to make the phase transition, one has to overcome the segment initial distortion by raising the pressure from $P_{C0}$ to $P_C$ at the same $T_{Cx}$. For the solution, the $\Delta E_x$ varies with the solute type and solute concentration, because the solute-solute interaction changes with solute concentration and type [83].

This principle of multifield mediating $P_C$ or $T_C$ for phase transition also applies to the phase transition of the nanosized water under compression since the molecular undercoordination has the same functionality of mediating the $\Delta E_x$. Mechanical compression shortens the O:H nonbond ($\beta_L(p) = -dd_L/Ldp > 0$) and lengthens the H−O bond ($\beta_H(p) < 0$) [52], which derives the segmental deformation energies of salt solution, the $\Delta E_L$ loss or the $\Delta E_H$ gain governs the $\Delta P_C = P_C - P_{C0}$ for the phase transition of the solution.

10.2  NaX/H$_2$O solution compression icing: Hofmeister series

High-pressure Raman examination of the Liquid–VI and the VI–VII phase transition for the solute type resolved NaX/H$_2$O (X = F, Cl, Br, I) and concentrated Na/H$_2$O solutions confirmed consistently the aforementioned predictions [68]. The phonon spectra show the abrupt $P_{Cx}$ change for all the phase transitions, refer to Fig 5 for pure water instance.

Fig 10 shows the concentration dependence of the $P_{Cx}$ for the Liquid-VI and VI-VII transition for the concentrated NaI/H$_2$O solutions at 298 K. Results show consistently that compression shortens the O:H nonbond and stiffens its phonons but the H–O bond responds to pressure contrastingly throughout the course unless at the phased boundaries. At higher concentrations, say 0.05 and 0.10, the skin 3450 cm$^{-1}$ mode are more active in responding to pressure, which evidence the preferential skin occupancy of the I$^-$ anions that enhances the local electric filed. The $P_{Cx}$ abrupcy at phase boundaries shows the weakening of the O–O repulsion due to structure transition from one phase to the other.

The sharp $P_C$ fall at the phase boundary indicates that the structural transition weakens the O—O repulsion, see Fig 11a. The phonon spectra for the room-temperature Liquid-VI and then VI–VII transition of salt solutions unveiled the following [68]:

1) There are three pressure zones that correspond to the Liquid, Ice VI, and Ice VII phase, toward phase X [38].

2) Mechanical compression recovers the energy storage by ionic solvation as both relax the O:H–O bond length and energy contrastingly. Polarization shortens the H–O bond and lengthens the O:H nonbond. The pressure stiffens the O:H phonon and softens the H-O phonon monotonically except for situations at the phase boundaries.

3) At phase boundaries, the measured pressure drops as the geometric restructuring that weakens the O—O repulsion. The weaker O—O makes both the O:H and the H–O contract abruptly under compression.

4) The $P_{Cx}$ increases with the anion radius difference between Na and X, following the I > Br > Cl > F ≈ 0 Hofmeister series order.

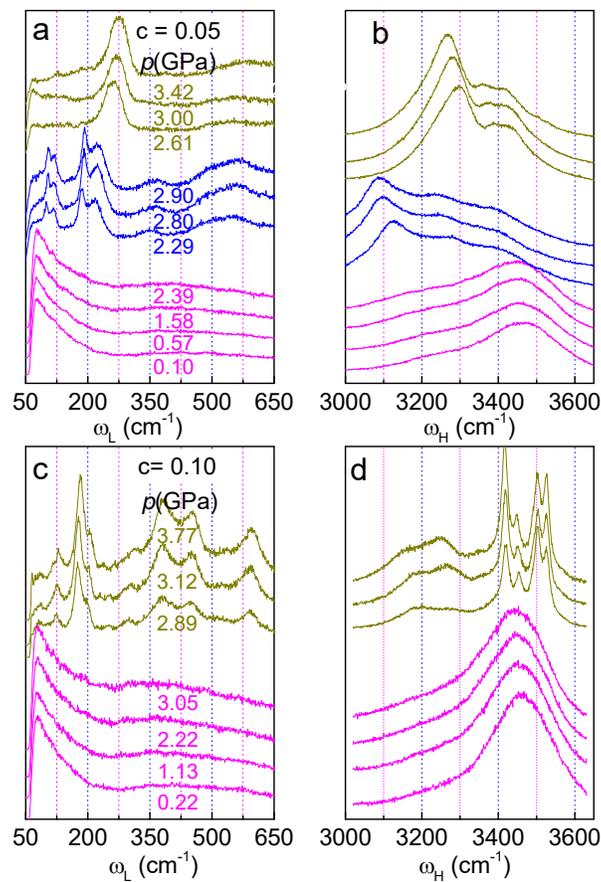

**Fig 10 | Compression and solute concentration resolved Liquid-VI-VII phase transition.** Spectral transition from the (a, b) Liquid-VI and VI-VII series to the (c, d) Liquid-VI-VII triple phase junction when the $NaI/H_2O$ concentration reaches 0.1 molar ratio (Reprinted with permission from [89])

10.3    Concentrated $NaI/H_2O$ icing: interanion repulsion

Fig 11b shows the NaI/H$_2$O concentration dependent P$_{Cx}$ for the consecutive Liquid–VI and VI–VII transition at 298 K. The trends of phonon frequency change hold the same to that of the compressed NaX/H$_2$O solutions. At higher solute concentrations, 0.05 and 0.10 for instance, the skin mode at 3450 cm$^{-1}$ responds more actively to the compression. This observation shows the I$^-$ preferential skin occupancy that strengthens the local electric filed. The abnormal P$_{Cx}$ sharp fall at phase boundaries results from the weakening of the O—O repulsion when transiting from one structure phase to another. The compression-resolved Raman spectra for the concentrated NaI/H$_2$O solutions unveiled the following, see Fig 11b:

1) The P$_{C1}$ for the Liquid-VI phase transition increases quickly than the P$_{C2}$ for the VI-VII transition with the increase of the NaI concentration. The P$_{C1}$ moves toward the P$_{C2}$ and eventually meets at 3.3 GPa and 350 K triple phase junction.
2) The P$_{C2}$ changes slowly with solute concentration and follows the VI–VII boundary in the phase diagram, which is in contrasting to the trend of solution type variation, in Fig 11a.
3) The concentration trend of the P$_{C1}$ moves along the Liquid–VI phase boundary, which is equivalent to the simultaneous compressing and heating in the phase diagram, see the phase diagram.
4) The solute concentration-resolved P$_{C1}$ and P$_{C2}$ trends indicate the effect of ionic electric field that varies with ionic concentration.

The involvement of anion-anion repulsion results in the P$_{Cx}$ inconsistency between the solute-type and the solute-concentration resolved solutions. The inter-anionic repulsive interaction weakens the electric field of the hydration volume, particularly at higher solute concentrations [83]. The polarization of water molecules inside the hydration volume screens the field of the Na$^+$ and thus the Na$^+$ holds a constant field strength within its hydration volume. The cation's hydration volume and its field strength are insensitive to the presence and concentration of other solutes. So, one can focus on the performance of the anionic solutes in the solutions. The solute concentration increase modulates the local electric field and the extent of the initial H–O bond energy storage. Thus, different solute types at a constant concentration resolve their field strengths inside the anionic hydration volume. Therefore, the P$_{Cx}$ changes with solute type, which follows the Hofmeister series order.

In contrast, the P$_{C2}$ is less sensitive than the P$_{C1}$ to the change of concentration of the same type of solute. The fast P$_{C1}$ growth with the increase of the I$^-$ shows the effect of the inter-anionic interaction that weakens the local electric field and the extent of polarization. The weakening of the polarization makes the O:H–O bond be easily compressed, and hence the $P_{C1}$ grows with concentration. However, the highly-compressed O:H–O bond is less sensitive to the local electric field of the hydration volume and the compression, which makes the $P_{C2}$ stable. The deformed O:H–O bond is harder to be deformed further than those less-deformed under the same pressure.

5)

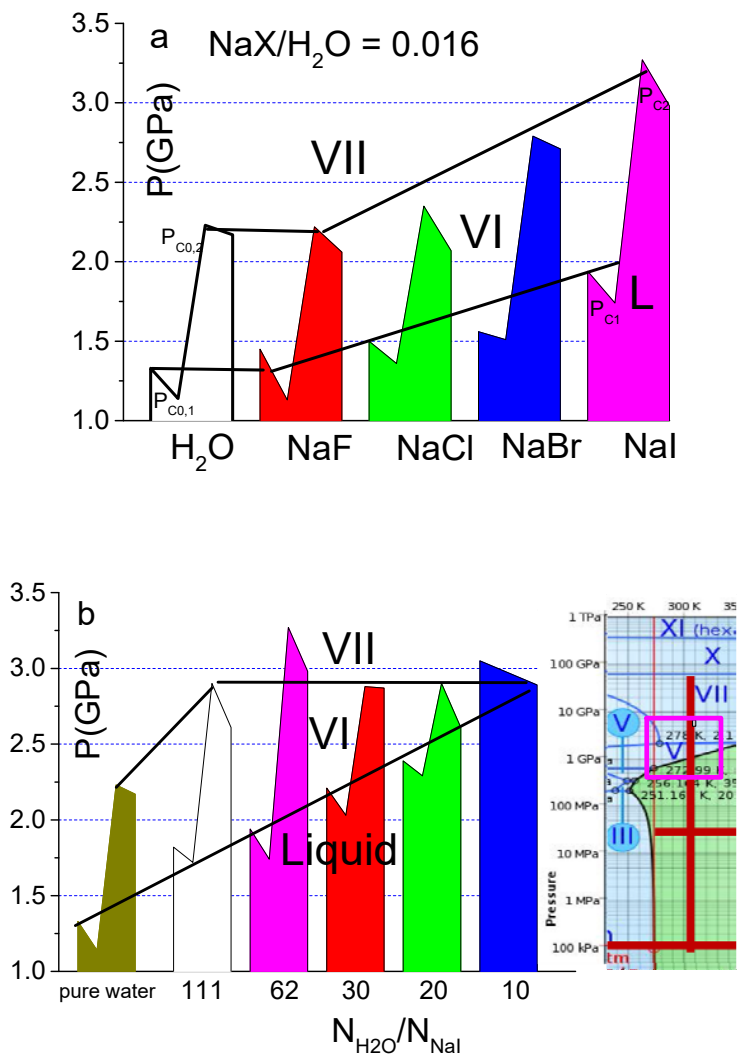

**Fig 11 | Solute type and concentration resolved $P_{Cx}$ for the Liquid–VI and the VI–VII phase transition of the 298 K salt solutions** [68, 89]. (a) NaX/H₂O type (0.016 molar ratio) and (b) NaI/H₂O concentration resolved $P_{Cx}$ for the ice formation. Inset b framed the processes of Liquid-VI-VII phase transition (reprinted with permission from [68, 89])

## 11  Icing by dynamic impulsion and static immersion of electrolyte

*− Mechanical impulsion and salt solvation discriminate the $T_C$ via O:H−O relaxation*

It is amazing that water freezes more likely when the fast-moving catalyst impinges upon the air−water interface and then get into the liquid. Such catalyst impact freezing has relevance for ice initiation in Earth's atmosphere and the snow Ice-QS transition by salting [90]. If a soluble substance is dissolved in water, both the $T_m$ and $T_N$ are intuitively reduced. Research [91] showed that the impinge of clay, sand, and silver iodide (AgI) could trigger freezing at a higher temperature in the contact mode (intrude with a momentum) than in the immersion mode (immerse the salt into the solution without initial momentum). In contrast, salt and sugar contact could initiate the freezing at 262 and 259.5 °K, respectively.

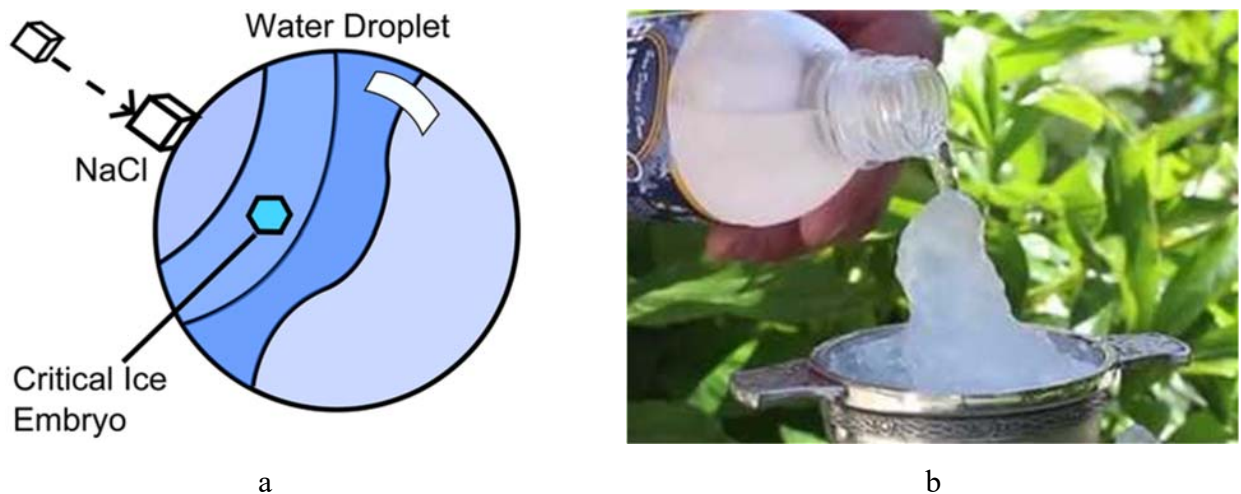

a                                                                                       b

**Fig 12. | Supercooled water freezing by mechanical and flying salt particle impulsion.** Schematic illustration of (a) flying salt (NaCl) collision with critical ice embryo [90] and (b) mechanical impulsion raised critical temperature for homogeneous and eutectic freezing of supercooled water.(reprinted with permission from [90])

Niehaus and Cantrell [90] demonstrated that (K, Na)(Cl, I) and (Na, K)OH collision with the moderately supercooled water triggers freezing within 10 ns time scale at temperatures higher than 239 °K for the homogeneous ice nucleation of supercooled water. Detailed examination of the effect of particle size and density, impact velocity (~3 m/s), and collision kinetic energy suggested that the freezing behavior depends on the collision itself. The catalysts only lower the $T_m$ upon their dissolving into the liquid, as illustrated in Fig 12. Alternatively, the contact mode of solute freezing ls less effective than the immersion mode. The kinetic energy from a mechanical contact impulse is the key factor in reducing the energy barrier for nucleation, enhancing the probability of the QS-Ice phase transition.

Known mechanisms for the collision modulation of the $T_N$ include the following:

1) Subcritical ice embryos were adsorbed to the surface of the incoming particles, which serves as seeds to promote freezing [92];
2) The heat of wetting reduced the moment of coming particles, which lowers the free-energy barrier between water and ice [93]; and,
3) The free energy barrier at a three phase contact line is intrinsically reduced by collision [94, 95].
4) Salt endothermic salvation raises the $T_N$ [96].

Knollenberg [96] proposed that the endothermic solvation of salt solute modulates the collision ice nucleation. Salt impinging into water cools down its surrounding liquid as it absorbs heat to break the ionic bonds and to form hydrate volumes. If the water temperature is lower than the $T_{eutectic}$ for the water-salt system, freezing is possible. Alternatively, water may be cooled below its $T_N$ by the endothermic solvation of salt, freezing takes place before the ions being diffused into the supercooled region.

According to the presented premise of multifield phase transition determined by the segmental specific-heat disparity, the salt collision and immersion induced freezing or melting can be decomposed into the following factors:

1) Mechanical impulsion supplies a short pulse of pressure, $\Delta P = mv/(s\Delta t)$, that raises the $T_N$ and lowers the $T_m$ [22], with m, v, s, and t being the mass, velocity, area and the time span of the flying object impact on water.
2) Solvation injects ions that polarize the O:H−O bond to lower the $T_N$ and raise the $T_m$ [97].
3) The supersolid skin has the same effect of polarization with lowered $T_N$ and raised $T_m$ [98].

4)  The impinge of the charged solute involved two sequential processes of compression and solvation.

Mechanical collision effects the same to a momently compression, which is opposite of ionic polarization or molecular undercoordination in relaxing the O:H−O bond segmental length and energy. The $T_N$ for the QS-Ice transition depends on the O:H bond energy. The $T_m$ for QS-Liquid transition depends the H−O energy. The $T_m$ and $T_N$ are correlated with the interplay of the specific-heat and with the operating conditions that relaxes the segmental energy and vibration frequency. Therefore, the sequential events of impulsion modulate the $T_N$ and $T_m$ oppositely−collision raises but ionic solvation lowers the $T_N$:

1) First, collision of the flying salt particle with the supercooled QS droplet provides an impulse that promotes local QS−Ice transformation. This transition is extremely sensitive to the extent of mechanical perturbation. The typical situation is the instant ice formation by slapping the bottle of supercooled water, see Fig 12b.
2) Secondly, when the salt particle is dissolved, ionic polarization lowers the $T_N$ and raises the $T_m$. This process is solute type and concentration dependence. The eutectic temperature undergoes elevation upon collision and then depression upon solvation. The immersion of salt will only raise the $T_N$ and lowers the $T_m$, instead.
3) The order of mechanical impulsion and ion polarization matters the $T_m$ and $T_N$ differently by relaxing the O:H−O segment.

## 12    Dimension, pressure, and temperature compensated phase transition

*− Heating compensates compression in Ice–QS transition while undercoordination maximizes QS*

### 12.1    Nanosized Ice-QS transition by heating and compression

Atomic force microscopy examination [99] revealed the reversible transition from two-dimensional (2D) ice into a QS phase confined between the hydrophobic graphene and muscovite mica by mechanical compression at the ambient temperature. This discovery confirmed the joint effect of compression, molecular undercoordination, and substrate heating on the QS boundary dispersion. Compression disperses the QS boundary inwardly but the undercoordination outwardly. Fig 13 shows two facts as a consequence of compression-undercoordination- heating on the $P_C$ and $T_N$ for the Ice-QS transition or homogeneous ice formation [67]:

1) The $P_C$ is much higher for low-dimensional phase transition. At room temperature, the $P_C$ for the confined Ice–QS transition amounts at 6.0 GPa that is much higher than the $P_C$ at 1.3 or 3.5 GPa for the Liquid–Solid transition of pure water or 0.1 molar concentrated $NaI/H_2O$ solution [68, 89].
2) The $P_C$ the substrate temperature $T_S$ compensates each other. The $P_C$ drops as the $T_S$ increases.
3) The ice and QS coexist in the temperature range between 293 and 333 K, which is higher than the bulk QS phase existing between 258 and 277 K at the same ambient pressure [53].

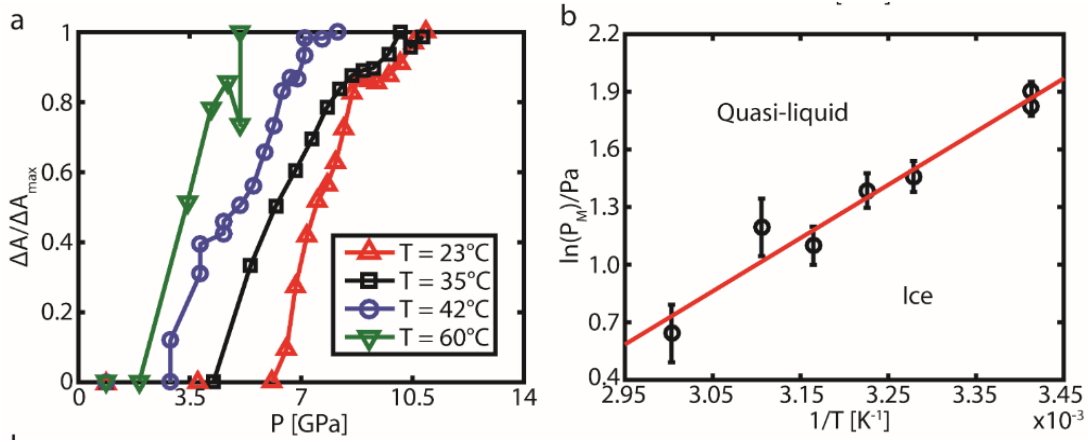

**Fig 13 | Phase diagram for the 2D QS-Ice phase transition.** (a) Ice patch area variation (QS–Ice transition) as a function of pressure and substrate temperature and (b) compensation of heating and compression on the Ice-QS (or quasi-liquid) phase boundary (reprinted with copyright permission from [99])

Molecular undercoordination raises the $T_m$ and lowers the $T_N$ by outward dispersing the QS boundaries. The $T_m$ for a monolayer [100] and the skin of bulk water [98] go up to 325 K and 310 K, respectively. Water inside single-walled carbon nanotubes (SWCNT) shows $T_m$ elevation by as much as 100 °K [101] and the $T_N$ for water within the SWCNT of 1.15 nm diameter is between 238 and 284 °K. The $T_N$ for a 4.4, 3.4, 1.4 and 1.2 nm sized droplet drops from 258 °K to 242 [102], 220 [102], 205 [69] and 172 °K [103], respectively. For a cluster of 18 molecules or less, the $T_N$ is about 120 °K [104]. These observations evidenced that molecular undercoordination raises the $T_m$ and lowers the $T_N$ and by QS boundary dispersion. These facts explain the coexistence of the confined Ice–QS phase between 299 and 333 °K [99], instead of 258 and 277 °K for bulk water.

The confinement has the same effect of ionic polarization that lengthens and softens the O:H nonbond. Compression does it contrastingly to raise the $T_N$ for translating the room-temperature confined ice into the QS phase at 299 K and below [99]. Compression up to 6 GPa raises the $T_N$ from far below to the room-temperature of confined ice. On the other hand, within the QS phase, the H–O bond follows the regular rule of thermal expansion and heating shortens the O:H nonbond, which eases the O:H–O compression. Therefore, QS heating lowers the $P_C$ for the Ice-QS transition, see Fig 13. Heating and compression compensate each other on O:H–O relaxation in the QS phase.

## 12.2 QS droplet compressibility

Fig 14 shows the compensation of P and T on the compressibility, $\kappa_T$, for micrometer sized water droplets in the 180–310 °K temperature and 1–1700 bar (1 bar = $10^5$ Pa) pressure ranges, which crosses the Ice–QS–Liquid phases [105]. The solid lines from right to left show the pressure increase. The P and T compensate each other to keep the $\kappa_T$ a constant, which means that one wants to have the same $\kappa_T$ at a higher pressure he must lower the temperature. The situation is the same as presented in Fig 13 b where the phase transition from ice to QS at the boundary is equivalent to the present $\kappa_T$. Consistency between the Fig 13 b and Fig 14 indicates that the supercooling QS presents. In the QS phase, the H–O bond undergoes cooling contraction and the O:H nonbond elongation, but compression always does them contrastingly. Therefore, so the mechanical compression and thermally heating compensate each other to keep the $\kappa_T$ constant. In contrast, heating shortens but compression always lengthens the H−O bond and the heating and compression could compensate each other in liquid or ice $I_h$ and $I_c$ phase [52, 53]. The higher temperature, the higher pressure is required to maintain the $T_C$ or the compressibility of the ice I or liquid.

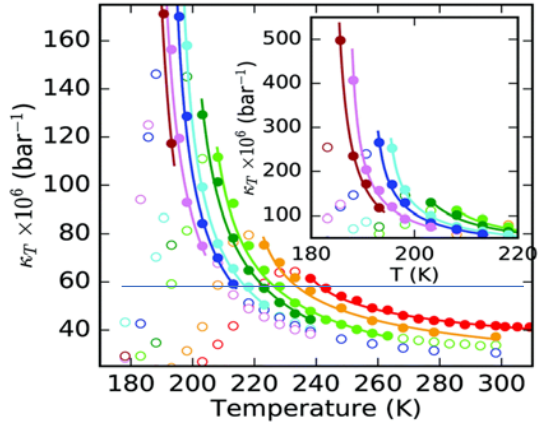

**Fig 14 | Temperature dependent compressibility (scattered symbols), $\kappa_T$, and its T–P compensation of microstructure liquid** [105]**.** Power-law simulating to pressures (from r to l: 1, 500, 1000, 1200, 1400, 1500, 1600 and 1700 bar). The inset shows $\kappa_T$ at the lower temperature range. (Reprinted with permission from [105])

13    Polarization: bandgap and dielectrics

− *Lone pair polarization widens the bandgap and raises the dielectric permittivity*

**Fig 15**a shows the band gap expansion from 4.5 to 6.6 eV when the pressure rises to 60 GPa. Density functional theory (DFT) calculations [52] derived the evolution of the density-of-states (DOS) covering the conduction and the valence bands for the compressed ice-VIII. The bottom edge of the valence band shifts from -6.7 eV at 1.0 GPa to -9.2 eV at 60 GPa, but its upper edge at the Fermi level stays unchanged. The band above $E_F$ shifts from 5.0 – 12.7 eV at 1 GPa to 7.4 – 15.0 eV at 60 GPa. Compression pushes the DOS above the $E_f$ further to expand the band gap, as shown in Fig 15b. Fig 15c shows a collection of the UV absorption spectra of the compressed ice [106]. The onset of UV absorption corresponds to the bandgap evolution, shifts positively with increasing pressure. Bandgap expansion makes ice more blueish and transparent as inset a showed.

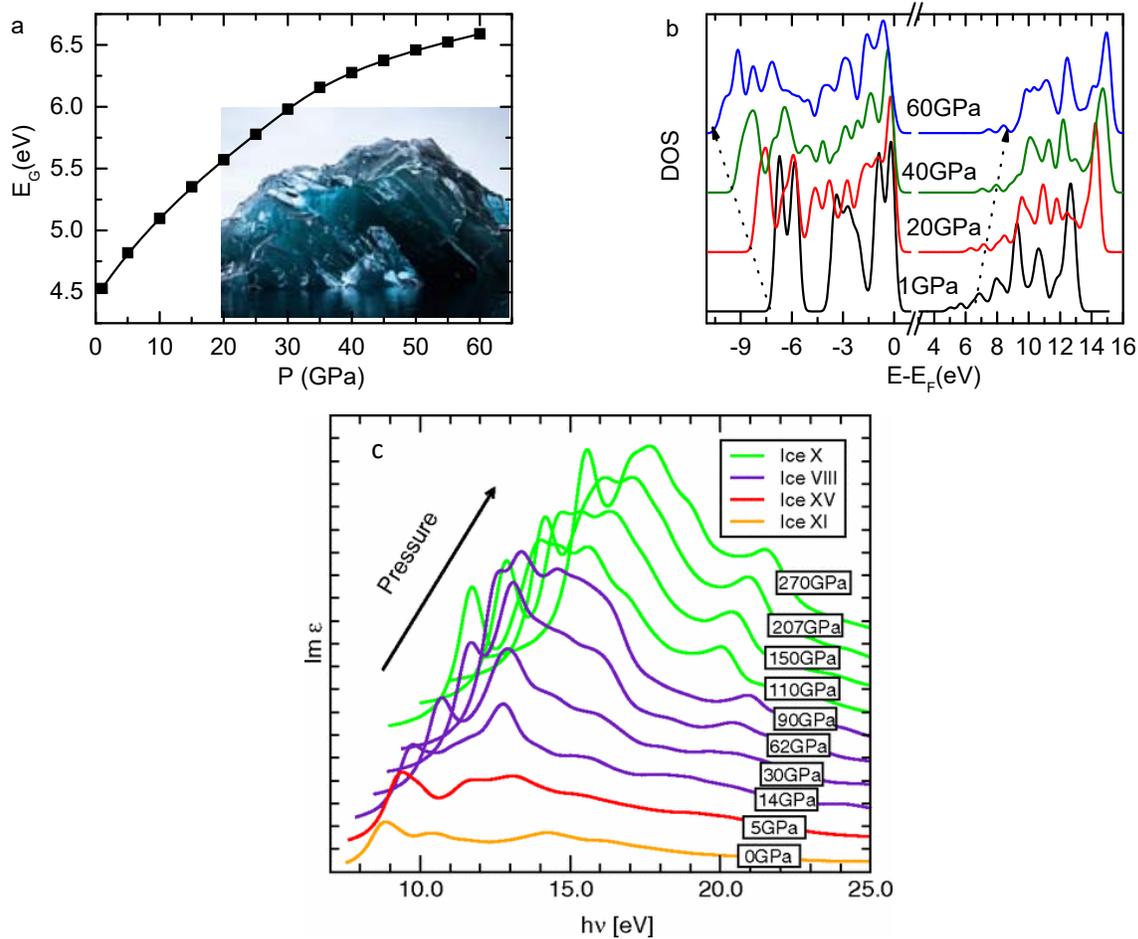

**Fig 15 | Bandgap expansion by nonbonding electron polarization of the compressed ice-VIII** [52]. Compression widens (a) the bandgap by (b) reserving the upper edge of the valence band and polarizing the lone-pair DOS, while entrapping the valence DOS bottom edge. Compression offsets (c) the band edge optical absorption of the compressed ice [106]. Inset a shows a upside-down iceberg that reflects blue light (photo credit: Alex Cornell, 2015 [28])(reprinted with permission from [52, 106])

Generally, the cohesive energy of the representative bond determines the bandgap of a semiconductor with involvement of electron–phonon coupling interactions [107-109]. Mechanical compression shortens and stiffens the representative bond and hence enlarges the bandgap [108]. Amazingly, bandgap expansion of ice follows the same pressure trend of "normal" semiconductors or insulators, but it arises from a completely different mechanism, because compression weakens the H−O bond, instead. Compression stiffens the weaker O:H bond whose contribution is insignificant. Therefore, the bandgap expansion of the compressed ice arises not from the $E_H$ reduction or the $E_L$ increase [52]. Compression deepens the bottom edge of the valence band and densifies the valence DOS, see Fig 15b. The densely entrapped valence DOS polarizes the lone-pairs to expand the bandgap [110]. The pressure resolved upward shift of the UV absorption also shows the strong polarization of the lone-pair states.

Electron transition from the valence band to the conduction band gives rise to the static dielectric constant of a semiconductor. Processes of bandgap $E_G$ width, lattice relaxation, and electron–phonon coupling determine the dielectric relaxation [107]. The dielectric permittivity ($\chi = \varepsilon_r - 1$) of a semiconductor is approximately proportional to the inverse square of its bandgap, $\chi \propto E_G^{-2}$ [111, 112]. The $\chi$ is therefore inversely proportional to the bond energy. The refractive index, $n = \varepsilon_r^{1/2} = (\chi + 1)^{1/2}$, drops accordingly when the specimen is compressed or reduces its size [107]. The dielectric constants of a semiconductor is lower in the skin region because of the atomic undercoordination derived bond contraction and bond strength gain [107].

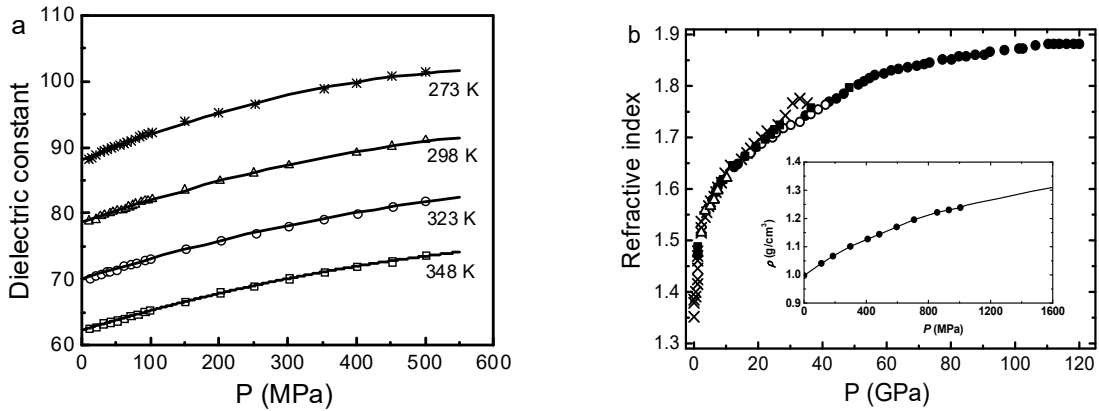

**Fig 16 │Water dielectric constant elevation by compression** [26, 27]. (a) Compression and liquid cooling raise the dielectric constant compared with (b) pressure enhanced refractive index and mass density (inset) of the ambient water. (reprinted with permission from [26, 27])

Compression or liquid cooling lengthen and weaken the H−O bond, which should conventionally narrow the bandgap and raise the dielectric permittivity, but the bandgap did the otherwise. For water ice, the dielectric constant increases with pressure and with the drop of temperature. The refractive index of liquid water at room temperature follows the same trend of mass density, see Fig 16. These observations disobey the rules for bandgap – dielectric correlation for semiconductors. H−O bond softening should narrow the bandgap though it raises the dielectric constant.

The abnormal bandgap – dielectric correlation of the compressed water ice may follow the possible mechanisms, but the polarized lone-pair density is most favorable,

$$\chi \propto \begin{cases} \rho_m(d_{OO}) & (Mass\ density) \\ E_G^{-2} & (Bandgap) \\ \rho_e(polarization) & (Electron\ density) \end{cases}$$

Compression shortens the O—O distance and raises the mass density but lowers the $E_H$. Polarization densifies the lone pairs by increasing its saturation and raises the energy of the nonbonding lone pairs. Water molecular polarization happens under compression, cooling, and molecular undercoordination, which should increase both bandgap and the dielectric permittivity. Observations of bandgap expansion and the dielectric permittivity elevation favor consistently the dominance of lone-pair polarization of the compressed water ice.

## 14 Compressibility, viscosity, and thermal diffusivity

*− O—O repulsion minimizes the compressibility and polarization governs the viscosity*

Liquid water has a low compressibility and reaches its lowest at temperatures around 320 K [30]. Fig 17 shows that the compressibility of liquid water is lower than that of ice. The compressibility reduces from 5.1 to 4.4 × 10$^{-10}$ Pa$^{-1}$ at the zero-pressure limit when heating from 273 °K to temperatures around 315 °K. A 100 MPa pressure lowers the compressibility to 3.9×10$^{-10}$ Pa$^{-1}$ at 273 °K [113]. The low compressibility of water means that even in the deep oceans of 4 km depth, where the pressure reaches 40 MPa, there is only a 1.8% decrease in volume and 0.6% compression of the O:H−O bond [114].

The thermal diffusivity, $\alpha = \kappa/(\rho C_P)$, arising from lattice vibrations and it depends on the thermal conductivity $\kappa$, mass density ρ, and specific heat capacity at constant pressure $C_P$. The α increases at the ambient temperatures to a maximum at about 0.8 GPa [115]. Compression lowers the viscosity of water at temperatures below about 300 °K under 100 – 200 MPa pressure. The viscosity passes through a pressure minimum and then increases [116]. In contrast, compression of most other liquids leads to a progressive loss of fluidity as molecules are squeezed closer together [29].

The pressure effect on the thermal diffusivity, viscosity, and compressibility is very complicated and no assertion has been reached. The pressure-viscosity behavior is explained by the balance between hydrogen-bonding network cohesive strength and the van der Waals force (called dispersion) [117]. The viscosity recovers at higher temperature, because heating breaks the balance, being in favor of the dispersion forces [118]. The viscosity minima at different temperatures is explained as the conversion of the low-density state to the high-density water under pressure [117].

From the bonding and electronic dynamics viewpoint, factors determine the pressure-related properties include the compression-induced polarization, O:H and O—O contraction, H−O elongation, O—O repulsion that varies with the interionic separation and the extent of polarization, and the response of these factors to heating in different phases or temperature ranges. The actual situation is much more complicated for a model because the O:H−O is extremely sensitive to a perturbation.

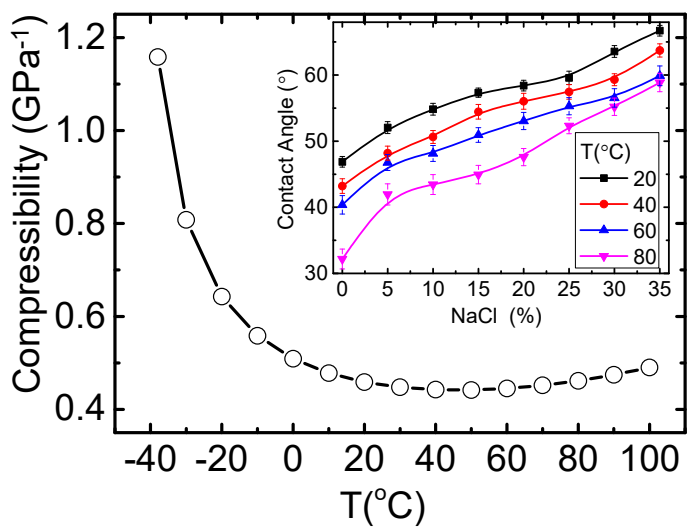

**Fig 17 | Temperature mediated compressibility and polarizability** (inset). Thermal depolarization recovers the least compressibility at around 323 °K (Reprinted with permission from [119].)

Compression shortens the O—O distance and the interionic repulsion opposes the compression, which lowers compressibility of water ice compared to normal materials [52]. Compression at sufficiently low temperature depresses the viscosity of water as compression elongates and softens the H−O bond. The temperature resolved compressibility displays the joint effect of compression and heating. Compression shortens the O:H−O bond and enhances the polarization but heating does the opposite. The inset in Fig 17 shows that salting enhances but heating depresses the polarization that provides repulsive force between oxygen ions. For instance, the contact angle drops from 320 to 303 °K when the droplet is heated from 293 to 353 °K. The weakening of interionic repulsion and thermal depolarization eases the recovery of the compressibility as temperature rises. Therefore, opposing compression, polarization takes the responsibility for the compressibility minimum around 320 °K [97]. When heated, depolarization proceeds, which raises the compressibility slightly. When cooled below the least compressibility temperature, mechanical compression and cooling contraction of the O:H nonbond work cooperatively, elevation the compressibility.

## 15 Summary

− *The premise of bonding and electronic dynamics leads to the core science of water ice*

The premise of hydrogen bonding and electronic dynamics has enabled the analytical forms of the dQ/(Qd$q$) coefficients for multifield (q = P, T, z, ε) mediated properties (Q = d, ω, E, T$_C$). From this perspective, we have been able to substantiate the understanding of anomalies demonstrated by water ice upon compression and its joint effect of heating, polarization and molecular undercoordination. Proper choice of H$^+$ as the O:H−O coordination origin, taking the O—O repulsive coupling and polarization into action, and introducing the segmental specific-heat make great differences. We may summarize the progress as follows:

1) The O:H−O segmental cooperativity and its specific-heat disparity dictate the extraordinary adaptivity, recoverability, and sensitivity of water ice in responding to compression and other physical perturbations.

2) Compression shortens the O:H nonbond and stiffens its phonon while the H−O responds to compression contrastingly unless at grain boundaries or under extreme conditions, which disperses the QS boundary inwardly through Einstein's relation, resulting in ice regelation ($T_m$ depression) and instant ice formation ($T_N$ elevation or depression). Ice regelation shows the strong recoverability of O:H−O deformation and dissociation.

3) Lagrangian-Laplacian resolution to the coupled O:H−O oscillation dynamics transforms the measured ($d_x$, $\omega_x$) into the ($k_x$, $E_x$), turning out the potential paths of the segmented O:H−O under compression.

4) Reproduction of the phase boundaries clarifies that: (i) H−O bond relaxation dictates transition at $dT_C/dP < 0$ phase boundaries, for $T_m$ and VII−VIII boundary instances; (ii) O:H relaxation governs the $dT_C/dP > 0$ boundaries, for $T_N$ and $T_V$ instances; (iii) both O:H and H−O frozen at $dT_C/dP = 0$ boundaries (parallel to P-axis) at ultra-low temperatures (for $I_C$−XI transition); (iv) O:H and H−O energetic competent relaxation dictates $dT_C/dP = \delta(P_C)$ boundaries (parallel to the P−axis, for VIII−X transition instance).

5) Compression widens the bandgap and raises the dielectric constant of water ice by polarization rather than arising from H−O bond energy that happens to other regular semiconductors.

6) The competition among the O—O repulsion and compression and the lone-pair polarization minimizes the compressibility of water to a minimum at 313 °K; O:H−O bond thermal depolarization and liquid O:H thermal expansion recovers its compressibility. The combination of lone-pair polarization and O:H−O relaxation determines the viscosity and thermal diffusivity though it is subject to further varication.

Progress showed the efficiency and essentiality of an alternative way of thinking about the O:H−O bonding and electronic dynamics. Further extension of this scheme would provide impact to engineering molecular crystals with lone-pair involvement such as drugs, explosives, foods, and processes of water-protein interactions.

Declaration

No conflicting interest is declared

Acknowledgement

Financial support received from National Natural Science Foundation (Nos. 21875024;11872052) of China is acknowledged.
References

1. M. Faraday, *Note on Regelation.* Proceedings of the Royal Society of London, **10**: 440-450, (1859).
2. J. Thomson, *Note on Professor Faraday's Recent Experiments on Regelation.* Proceedings of the Royal Society of London, **11**: 198-204, (1860).
3. T. Hynninen, V. Heinonen, C.L. Dias, M. Karttunen, A.S. Foster, and T. Ala-Nissila, *Cutting Ice: Nanowire Regelation.* Physical Review Letters, **105**(8), (2010).
4. D. Petely. *Our strange desire to find a landslide trigger*. http://ihrrblog.org/2013/11/08/our-strange-desire-to-find-a-landslide-trigger/ 2013.
5. X. Zhang, Y. Huang, P. Sun, X. Liu, Z. Ma, Y. Zhou, J. Zhou, W. Zheng, and C.Q. Sun, *Ice Regelation: Hydrogen-bond extraordinary recoverability and water quasisolid-phase-boundary dispersivity.* Sci Rep, **5**: 13655, (2015).
6. C.T. Calderon, *Premelting, Pressure Melting, and Regelation of Ice Revisited.* Journal of Applied Mathematics and Physics, **6**(11): 2181, (2018).
7. J.D. Goddard, *The viscous drag on solids moving through solids.* Aiche Journal, **60**(4): 1488-1498, (2014).
8. G. Malenkov, *Liquid water and ices: understanding the structure and physical properties.* Journal of Physics-Condensed Matter, **21**(28): 283101, (2009).
9. J. Chen, K. Nagashima, K.-i. Murata, and G. Sazaki, *Quasi-liquid layers can exist on polycrystalline ice thin films at a temperature significantly lower than on ice single crystals.* Crystal Growth & Design, **19**(1): 116-124, (2018).
10. K. Aoki, H. Yamawaki, and M. Sakashita, *Observation of Fano interference in high-pressure ice VII.* Physical Review Letters, **76**(5): 784-786, (1996).
11. M. Song, H. Yamawaki, H. Fujihisa, M. Sakashita, and K. Aoki, *Infrared absorption study of Fermi resonance and hydrogen-bond symmetrization of ice up to 141 GPa.* Physical Review B, **60**(18): 12644, (1999).
12. P. Pruzan, J.C. Chervin, E. Wolanin, B. Canny, M. Gauthier, and M. Hanfland, *Phase diagram of ice in the VII-VIII-X domain. Vibrational and structural data for strongly compressed ice VIII.* Journal of Raman Spectroscopy, **34**(7-8): 591-610, (2003).
44